\documentclass[review,12pt,authoryear]{elsarticle}

\usepackage{amssymb}
\usepackage{amsthm}
\usepackage{framed} 
\usepackage{amsmath,color}
\usepackage{mathrsfs}
\usepackage{graphicx}
\usepackage{epstopdf}
\usepackage{float}
\usepackage{caption}
\usepackage{subcaption}
\usepackage{bm}
\usepackage{bbm}
\usepackage{mathrsfs}
\usepackage{cleveref}
\usepackage{soul}
\usepackage{accents}
\usepackage{color,soul} 
\usepackage{color} 
\biboptions{sort&compress}
\soulregister\citep7 
\soulregister\citet7 
\soulregister\citealp7 
\newsavebox{\measurebox} 
\usepackage{titlesec} 
\usepackage[nomarkers,figuresonly]{endfloat} 
\usepackage[T1]{fontenc} 
\usepackage{lmodern} 
\input{glyphtounicode}  

\journal{Journal of the Mechanics and Physics of Solids}

\makeatletter
\def\@author#1{\g@addto@macro\elsauthors{\normalsize%
    \def\baselinestretch{1}%
    \upshape\authorsep#1\unskip\textsuperscript{%
      \ifx\@fnmark\@empty\else\unskip\sep\@fnmark\let\sep=,\fi
      \ifx\@corref\@empty\else\unskip\sep\@corref\let\sep=,\fi
      }%
    \def\authorsep{\unskip,\space}%
    \global\let\@fnmark\@empty
    \global\let\@corref\@empty  
    \global\let\sep\@empty}%
    \@eadauthor={#1}
}
\makeatother

\titleformat{\paragraph}
{\normalfont\normalsize\itshape}{\theparagraph}{1em}{}
\titlespacing*{\paragraph}
{0pt}{3.25ex plus 1ex minus .2ex}{1.5ex plus .2ex}

\begin{document}

\begin{frontmatter}



\title{The role of plastic strain gradients in the crack growth resistance of metals}


\author{Emilio Mart\'{\i}nez-Pa\~neda\corref{cor1}\fnref{Cam}}
\ead{mail@empaneda.com}

\author{Vikram S. Deshpande\fnref{Cam}}

\author{Christian F. Niordson\fnref{DTU}}

\author{Norman A. Fleck\fnref{Cam}}

\address[Cam]{Department of Engineering, Cambridge University, CB2 1PZ Cambridge, UK}

\address[DTU]{Department of Mechanical Engineering, Technical University of Denmark, DK-2800 Kgs. Lyngby, Denmark}

\cortext[cor1]{Corresponding author.}

\begin{abstract}
Crack advance from short or long pre-cracks is predicted by the progressive failure of a cohesive zone in a strain gradient, elasto-plastic solid. The presence of strain gradients leads to the existence of an elastic zone at the tip of a stationary crack, for both the long crack and the short crack cases. This is in sharp contrast with previous asymptotic analyses of gradient solids, where elastic strains were neglected. The presence of an elastic singularity at the crack tip generates stresses which are sufficiently high to activate quasi-cleavage. For the long crack case, crack growth resistance curves are predicted for a wide range of ratios of cohesive zone strength to yield strength. Remarkably, this feature of an elastic singularity is preserved for short cracks, leading to a severe reduction in tensile ductility. In qualitative terms, these predictions resemble those of discrete dislocation calculations, including the concept of a dislocation-free zone at the crack tip.
\end{abstract}

\begin{keyword}

Strain gradient plasticity \sep Length scales \sep Cohesive zone modelling \sep Finite element analysis \sep Fracture



\end{keyword}

\end{frontmatter}



\section{Introduction}
\label{Sec:Introduction}

Conventional plasticity theories, such as $J_2$ flow theory, predict that the tensile stress state ahead of a mode I crack in an elastic-perfectly plastic solid is on the order of three times the yield stress $\sigma_Y$. The factor of 3 arises from plastic constraint effects, and is explained in terms of the Prandtl stress field for a flat punch, see for example \citet{Rice1968a}. Ductile fracture by void growth is promoted by this high hydrostatic stress \citep{Rice1969,Hancock1976,McClintock1968}. This level of tensile stress is far below the cleavage strength (typically on the order of $10\sigma_Y$) yet cleavage fracture in the presence of significant plastic flow has been observed, see for example \citet{Elssner1994}, \citet{Bagchi1996}, \citet{Korn2002}. Additional physics is needed to explain the occurrence of cleavage failure in the presence of plasticity. As argued by \citet{Wei1997} and \citet{Jiang2010}, cleavage can occur if the stress elevation due to plastic strain gradients is sufficient to attain the ideal strength. This mechanism is reassessed in the present study.\\

The micromechanical basis for strain gradient effects is the elevation of dislocation-based hardening, and thereby of stress levels, as first appreciated by \citet{Nye1953} and \citet{Cottrell1964}. Additional, dislocation-based arguments were advanced by \citet{Ashby1970} and \citet{Brown1976}. In broad terms, plastic strain gradients demand the existence of geometrically necessary dislocations (GNDs), and this elevation in dislocation density increases the flow strength by mechanisms such as forest hardening \citep{Fleck1994}. Such strain gradient strengthening can explain a wide range of phenomena such as the Hall-Petch size effect, see for example \citet{Shu1999}. The Hall-Petch effect is based on grain-to-grain plastic anisotropy such that strain gradients are present at the grain-to-grain level when the macroscopic strain field is uniform (or non-uniform). Additional strain gradient effects arise when the macroscopic strain field is non-uniform, as near a crack tip, or in simple test geometries such as a wire in torsion \citep{Fleck1994}, a beam in bending \citep{Stolken1998} or at the tip of an indenter (\citealp{Stelmashenko1993}; \citealp{Poole1996}; \citealp{Nix1998}). A large literature has emerged on strain gradient plasticity (SGP) formulations (\citealp{Aifantis1984}; \citealp{Fleck1993,Fleck2001}; \citealp{Gao1999}). The pivotal step in constructing these phenomenological theories is to express the plastic work in terms of both plastic strain and plastic strain gradient, thereby introducing a length scale into the material description. Recent SGP models incorporate both dissipative (that is, unrecoverable) and energetic (that is, recoverable) gradient contributions (\citealp{Gudmundson2004,Gurtin2005}; \citealp{Fleck2009,Fleck2009b}).\\

Recently, the effect of strain gradients in elevating crack tip stress levels has been emphasized in a number of numerical investigations on stationary cracks \citep{Jiang2001,Komaragiri2008,IJSS2015,IJP2016}. It has been suggested that this elevation in stress influences fatigue damage (\citealp{Sevillano2001}; \citealp{Brinckmann2008}; \citealp{Pribe2019}), microvoid cracking \citep{Tvergaard2008}, and hydrogen embrittlement \citep{AM2016,IJHE2016}. Stress elevation due to plastic strain gradients is also relevant to propagating cracks. \citet{Wei1997}, and \citet{Wei2004}, quantified the dependence of steady state fracture toughness $K_{SS}$ upon material length scale $\ell$ for the \citet{Fleck1997} gradient theory and \citet{Gao1999} gradient theory, respectively. Recently, \citet{Seiler2016} computed the initial stages of the crack growth resistance curve for a viscoplastic, strain gradient plasticity theory \citep{Huang2004a}, and investigated the dependence of R-curves on viscoplastic constitutive parameters and on the intrinsic material length scale.\footnote{We note in passing that the \citet{Huang2004a} theory is a lower order theory that neglects higher order stresses. The present study assumes the presence of higher order stresses that are work conjugate to plastic strain gradients.} The recent strain gradient theory of \citet{Gudmundson2004} (see also \citealp{Gurtin2005,Fleck2009}) has additional features that can significantly influence crack growth resistance: this motivates the present paper. First, the recent asymptotic analysis of \citet{EJMaS2019} for a stationary crack in a dissipative strain gradient solid reveals the existence of an elastic crack tip zone, reminiscent of a dislocation-free zone \citep{Suo1993}. Second, both energetic and dissipative length scales enter the constitutive relations; their influence on fracture problems has not yet been assessed. Both features are explored here in the context of both stationary and propagating cracks. In addition, we explore the effect of crack length in relation to the material length scales and to the fracture length scale of the crack tip process zone. Thereby, insight is gained into the role of strain gradients on the behaviour of short cracks.\\

The present study is structured as follows. The constitutive model is presented in Section \ref{Sec:Theory}, including the phenomenological formulation of strain gradient plasticity, and the implicit finite element implementation. The asymptotic response at the tip of a stationary crack in a strain gradient solid is investigated in Section \ref{Sec:Stationary}. Crack growth is explored in Section \ref{Sec:Results} in two steps. First, R-curves are computed for a long crack by means of a cohesive zone, and the relative role of energetic versus dissipative strain gradient terms is quantified. Second, the short crack case is examined and we compute the sensitivity of the macroscopic stress versus strain response to crack length $a$ and to the material length scale $\ell$. The sensitivity of the tensile ductility to the presence of a short crack is emphasized: it is shown that strain gradients play a major role. Finally, concluding remarks are given in Section \ref{Sec:Concluding remarks}.

\section{Strain Gradient Plasticity}
\label{Sec:Theory}

\subsection{Flow theory}

\subsubsection{Variational principles and balance equations}

We adopt a small strain formulation. The total strain rate $\dot{\varepsilon}_{ij}$ is the symmetric part of the spatial gradient of the displacement rate $\dot{u}_i$, such that $\dot{\varepsilon}_{ij}=\left(\dot{u}_{i,j} + \dot{u}_{j,i} \right)/2$; $\dot{\varepsilon}_{ij}$ decomposes additively into an elastic part, $\dot{\varepsilon}^e_{ij}$, and a plastic part, $\dot{\varepsilon}^p_{ij}$. Write $\sigma_{ij}$ as the Cauchy stress, $q_{ij}$ as the so-called micro-stress tensor (work-conjugate to the plastic strain $\varepsilon_{ij}^p$) and $\tau_{ijk}$ as the higher order stress tensor (work-conjugate to the plastic strain gradient $\varepsilon_{ij,k}^p$). For a solid of volume $V$ and surface $S$, the principle of virtual work reads,
\begin{equation}\label{eq:PVW1Gud}
\int_V \Big( \sigma_{ij} \delta \varepsilon_{ij}^e + q_{ij} \delta \varepsilon_{ij}^p + \tau_{ijk} \delta \varepsilon_{ij,k}^p \Big) dV = \int_S \left( T_i \delta u_i + t_{ij} \delta \varepsilon_{ij}^p \right) dS
\end{equation}

The right-hand side of Eq. (\ref{eq:PVW1Gud}) includes both conventional tractions $T_i$ and higher order tractions $t_{ij}$. Write $\sigma_{ij}'$ as the deviatoric part of $\sigma_{ij}$, and write $n_k$ as the unit outward normal to the surface $S$. Then, upon making use of the Gauss divergence theorem, equilibrium within $V$ reads
\begin{align}
& \sigma_{ij,j}=0 \nonumber \\ 
& \tau_{ijk,k} + \sigma'_{ij} - q_{ij}=0 \label{eq:EQ2Gud}
\end{align}

\noindent and on $S$ reads,
\begin{align}
& T_i=\sigma_{ij} n_j \nonumber \\ 
& t_{ij}=\tau_{ijk} n_k
\end{align}

\subsubsection{Constitutive description}

\citet{Gudmundson2004} and \citet{Fleck2009} explain that both $q_{ij}$ and $\tau_{ijk}$ can have dissipative and energetic contributions: $q_{ij}=q_{ij}^D+q_{ij}^E$ and $\tau_{ijk}=\tau_{ijk}^D+\tau_{ijk}^E$, where the superscripts $D$ and $E$ denote dissipative and energetic, respectively. In general, the Cauchy stress $\sigma_{ij}$, along with $q_{ij}^E$ and $\tau_{ijk}^E$, are derived from the bulk free energy of the solid $\Psi$. In the present study, we shall assume that $q_{ij}^E$ vanishes and thus limit attention to a solid that displays isotropic hardening in the absence of a strain gradient. The significance of a finite value of $q_{ij}^E$ (with $\tau^D_{ijk} \equiv 0$) has been explored in the recent study of \citet{JAM2018}; here, we limit our focus to the role of kinematic hardening associated with the gradient of plastic strain. Accordingly, the bulk free energy $\Psi$ of the solid depends upon the elastic strain $\varepsilon_{ij}^e$ and the plastic strain gradient $\varepsilon_{ij,k}^p$ but not upon the plastic strain $\varepsilon_{ij}^p$, such that
\begin{equation}\label{Eq:FreeEnergy}
\Psi \left( \varepsilon_{ij}^e, \varepsilon_{ij,k}^p \right)=\frac{1}{2} \varepsilon_{ij}^e C_{ijkl} \varepsilon_{kl}^e + \frac{1}{2} \mu L_E^2 \varepsilon_{ij,k}^p \varepsilon_{ij,k}^p
\end{equation}

\noindent Here, $C_{ijkl}$ is the isotropic elastic stiffness tensor, $\mu$ is the shear modulus and $L_E$ is the energetic constitutive length parameter. Upon noting that
\begin{equation}
\delta \Psi = \sigma_{ij} \delta \varepsilon_{ij}^e + \tau_{ijk}^E \delta \varepsilon_{ij,k}^p
\end{equation}

\noindent the energetic stress quantities follow as
\begin{align}
& \sigma_{ij}=\frac{\partial \Psi}{\partial \varepsilon_{ij}^e}=C_{ijkl}\left( \varepsilon_{kl} - \varepsilon_{kl}^p \right) \label{eq:HookeLaw} \\
& \tau_{ijk}^E=\frac{\partial \Psi}{\partial \varepsilon_{ij,k}^p}=\mu L_E^2 \varepsilon_{ij,k}^p \label{eq:tauEGudmundson}
\end{align}

Now consider plastic dissipation. For both the rate dependent case, and the rate independent limit, we define the plastic work rate as
\begin{equation}
\dot{W}^p= \Sigma \dot{E}^p
\end{equation}

\noindent where $\Sigma$ is an effective stress, work-conjugate to a gradient-enhanced effective plastic strain rate $\dot{E}^p$. The latter is defined phenomenologically as
\begin{equation}\label{Eq:EpGud}
\dot{E}^p=\left( \frac{2}{3} \dot{\varepsilon}^p_{ij} \dot{\varepsilon}^p_{ij} + L_D^2 \dot{\varepsilon}^p_{ij,k} \dot{\varepsilon}^p_{ij,k} \right)^{1/2} 
\end{equation}

\noindent where $L_D$ is the dissipative length scale. Upon noting that
\begin{equation}
\delta \dot{W}^p = \Sigma \delta \dot{E}^p = q_{ij}^D \delta \dot{\varepsilon}^p_{ij} + \tau_{ijk}^D \delta \dot{\varepsilon}^p_{ij,k}
\end{equation}

\noindent the constitutive relations for the dissipative stress quantities read
\begin{equation}\label{Eq:qGud}
q_{ij}^D= \frac{2}{3} \frac{\Sigma}{\dot{E}^p}\dot{\varepsilon}^p_{ij} \,\,\,\,\,\, \textnormal{and} \,\,\,\,\,\, \tau_{ijk}^D=\frac{\Sigma}{\dot{E}^p} L_D^2 \dot{\varepsilon}^p_{ij,k}
\end{equation}

The effective stress $\Sigma$ is readily obtained by substitution of (\ref{Eq:qGud}) into (\ref{Eq:EpGud}) to give
\begin{equation}\label{Eq:EffectiveStress}
\Sigma=\left( \frac{3}{2} q_{ij}^D q_{ij}^D + \frac{1}{L_D^2} \tau^D_{ijk} \tau^D_{ijk} \right)^{1/2}
\end{equation}

\subsection{Numerical implementation}
\label{Sec:ABAQUS}

A robust and efficient finite element framework is now presented in order to model crack propagation in a rate independent gradient plasticity solid. An implicit time integration scheme is developed for both energetic and dissipative higher order contributions.\\

Gradient plasticity theories are commonly implemented within a rate-dependent setting, thereby taking advantage of computational advantages and circumventing complications in the corresponding time independent model associated with identifying active plastic zones (see, for example, \citealp{Nielsen2014}). The mathematical foundations and associated variational structure for both the rate dependent and rate independent cases are given by \citet{Fleck2009,Fleck2009b}. Here, we make use of the viscoplastic law by \citet{Panteghini2016}, and exploit the fact that it adequately approximates the rate-independent solution in a computationally efficient manner. The effective flow resistance is related to the gradient-enhanced effective plastic flow rate through a viscoplastic function,
\begin{equation}
\Sigma=\sigma_F \left( E^p \right) V(\dot{E}^p)
\end{equation}

\noindent where the current flow stress $\sigma_F$ depends upon an initial yield stress $\sigma_Y$ and on $E^p$ via a hardening law. Here, we adopt the following isotropic hardening law,
\begin{equation}\label{Eq:IsoHard}
\sigma_F=\sigma_Y \left( 1 + \frac{E E^p}{\sigma_Y} \right)^N
\end{equation}

\noindent in terms of the Young's modulus $E$ and strain hardening exponent $N$ $(0\leq N \leq 1)$. Following \citet{Panteghini2016} the viscoplatic function $V(\dot{E}^p)$ is defined as\footnote{This choice has the advantage that the consistent stiffness matrix, as defined in the Supplementary Material, remains finite as $\dot{E}^p \to 0$.}
\begin{equation}
V (\dot{E}^p) =   \begin{cases} 
   \dot{E}^p/ \left(2 \dot{\varepsilon}_0 \right) & \text{if } \dot{E}^p/\dot{\varepsilon}_0 \leq 1 \\
   1 - \dot{\varepsilon}_0 / \left( 2 \dot{E}^p \right)     & \text{if } \dot{E}^p / \dot{\varepsilon}_0 > 1
  \end{cases}
\end{equation}

\noindent in terms of a reference strain rate $\dot{\varepsilon}_0$. A sensitivity study for a sufficiently small choice of $\dot{\varepsilon}_0$ is conducted to ensure that the rate independent limit is attained in all the results presented subsequently. The reader is referred to \citet{Panteghini2016} for a more detailed interpretation of $\dot{\varepsilon}_0$.\\

The finite element scheme takes displacements and plastic strains as the primary kinematic variables, in accordance with the theoretical framework. $C_0$-continuous finite elements are adopted since the differential equations are of second order. The displacement field $u_i$ at position $\bm{x}$ is written in terms of the shape functions $N_i^n$ and associated nodal displacements $U^n$, where $n$ denotes the degree of freedom, such that
\begin{equation}
u_i=\sum_{n=1}^{D_u} N_i^{n} U^n
\end{equation}

\noindent Here, $D_u$ is the total number of degrees of freedom for the nodal displacements. Likewise, the plastic strain field $\varepsilon_{ij}^p$ is expressed in terms of the shape functions $M_{ij}^{n}$ and associated nodal quantities $\varepsilon_p^n$ as
\begin{equation}
\varepsilon_{ij}^p=\sum_{n=1}^{D_{\varepsilon_p}} M_{ij}^{n} \varepsilon_p^n
\end{equation}

\noindent where $D_{\varepsilon_p}$ denotes the total number of degrees of freedom for the nodal plastic strain components. Quadratic shape functions are employed for interpolation of both displacements and plastic strains. Accordingly, the plastic strain gradient $\varepsilon_{ij,k}^p$ and the total strain $\varepsilon_{ij}$ are related to the nodal plastic strains and displacements through $M_{ij,k}^n$ and the strain-displacement matrix $B_{ij}^n$, respectively; see the Supplementary Material for further details.\\

The non-linear system of equations is solved iteratively by the Newton-Raphson method from time step $t$ to $\left( t + \Delta t \right)$
\begin{equation}
\begin{bmatrix} \bm{u} \\ \bm{\varepsilon_p} \end{bmatrix}_{t+\Delta t} = \begin{bmatrix} \bm{u} \\ \bm{\varepsilon_p} \end{bmatrix}_{t} - \begin{bmatrix}
  \bm{K}_{u,u} & \bm{K}_{u,\varepsilon^p}\\
  \bm{K}_{\varepsilon^p,u} & \bm{K}_{\varepsilon^p,\varepsilon^p}
 \end{bmatrix}_t ^{-1}\begin{bmatrix} \bm{R}_u \\ \bm{R}_{\varepsilon^p} \end{bmatrix}_t
\end{equation}

\noindent where the residuals comprise the out-of-balance forces,
\begin{equation}
\bm{R}_u^n=\int_{V} \sigma_{ij} B_{ij}^n \, dV - \int_{S} T_i N_i^n \, dS
\end{equation}
\begin{equation}
\bm{R}_{\varepsilon^p}^n=\int_{V} \left[(q_{ij}-\sigma'_{ij})  M_{ij}^n  + \tau_{ijk} M_{ij,k}^n \right] dV - \int_{S} t_{ij} M_{ij,k}^n \, dS
\end{equation}

\noindent and the components of the consistent stiffness matrix $\bm{K}$ are obtained by differentiating the residuals with respect to the incremental nodal variables. Details are given in the Supplementary Material.\\

The numerical scheme is implemented in the commercial finite element package ABAQUS by means of a user element subroutine. To the best of the authors' knowledge, it constitutes the first Backward Euler implementation of the \citet{Gudmundson2004} class of strain gradient plasticity theories, including energetic and dissipative higher order contributions.\footnote{The code is made freely available at www.empaneda.com, hoping to facilitate research and enabling readers to reproduce the results.} The reader is referred to \citet{Danas2012c} and \citet{Dahlberg2013} for implicit implementations for the case of dissipative higher order stresses (with $\tau_{ijk}^E=0$).

\section{Stationary crack analysis}
\label{Sec:Stationary}

We assume that small scale yielding conditions prevail and we make use of a boundary layer formulation to prescribe a remote $K$ field. Consider a crack with its tip at the origin and with the crack plane along the negative axis of the Cartesian reference frame $(x,y)$. The elastic response of the solid is characterised by the Young's modulus $E$ and Poisson's ratio $\nu$. Then, an outer $K$ field is imposed by prescribing nodal displacements on the outer periphery of the mesh as
\begin{equation}
u_i = \frac{K}{E} r^{1/2} f_i \left( \theta, \nu \right)
\end{equation}
\noindent where the subscript index $i$ equals $x$ or $y$, and the functions $f_i \left( \theta, \nu \right)$ are given by
\begin{equation}
f_x = \frac{1+\nu}{\sqrt{2 \pi}} \left(3 - 4 \nu - \cos \theta \right) \, \cos \left(\frac{\theta}{2} \right)
\end{equation}
\noindent and
\begin{equation}
f_y = \frac{1+\nu}{\sqrt{2 \pi}} \left(3 - 4 \nu - \cos \theta \right) \, \sin \left(\frac{\theta}{2} \right)
\end{equation}

\noindent in terms of polar coordinates $(r, \theta)$ centred at the crack tip.
A representative value for the plastic zone size $R_p$ is given by the Irwin expression
\begin{equation}
R_p=\frac{1}{3 \pi} \left( \frac{K}{\sigma_Y}\right)^2
\end{equation}
\noindent for a stationary crack in an elastic, ideally plastic solid. Upon exploiting reflective symmetry about the crack plane, only half of the finite element model is analysed. A mesh sensitivity study reveals that the domain is adequately discretised by means of 5200 plane strain, quadratic, quadrilateral elements. The characteristic element size is $R_p/7500$ and the outer radius of the boundary layer is $5000R_p$, ensuring small scale yielding conditions.\\

\begin{figure}[H]
  \makebox[\textwidth][c]{\includegraphics[width=0.74\textwidth]{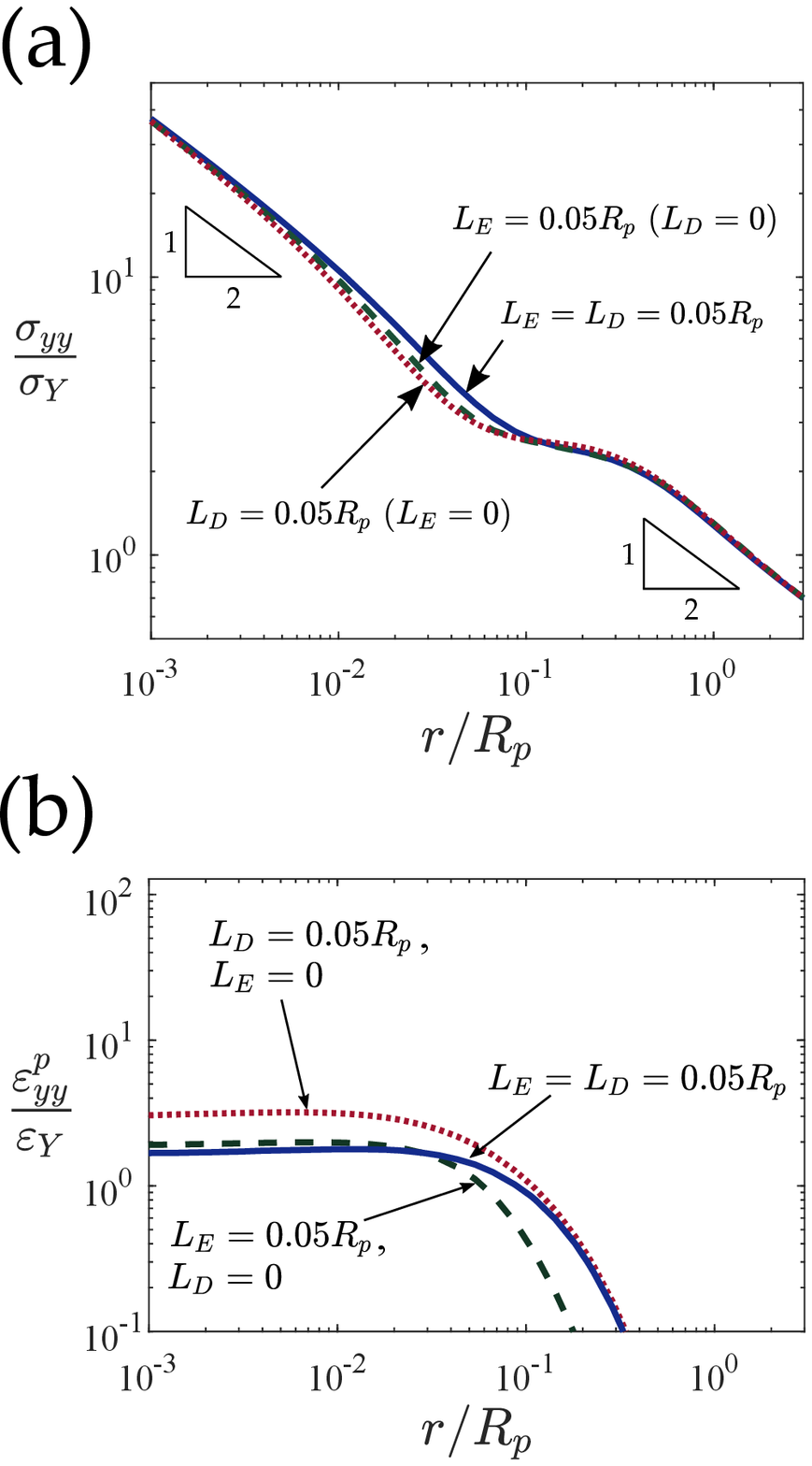}}%
  \caption{Finite element analysis of the asymptotic crack tip fields ($\theta=0^\circ$), (a) tensile stress component $\sigma_{yy}$, and (b) tensile plastic strain component $\varepsilon^p_{yy}$ for selected length scale parameters. Material properties: $\sigma_Y/E=0.003$, $N=0.1$, and $\nu=0.3$. Small scale yielding conditions.}
  \label{fig:CrackTipLeLd}
\end{figure}

A representative small scale yielding solution is presented in Fig. \ref{fig:CrackTipLeLd} for the choice $N=0.1$, $\sigma_Y/E=0.003$, and $\nu=0.3$. Insight is gained into the relative role of $L_E$ and $L_D$ by considering the three cases $L_E=L_D=0.05R_p$, $L_E=0.05R_p$ ($L_D=0$), and $L_D=0.05R_p$ ($L_E=0$). An elastic zone exists directly ahead of the crack tip if $L_E$ and/or $L_D$ is finite. The plastic strain $\varepsilon_{yy}^p$ reaches a plateau value over $0 < r < \ell$, see Fig. \ref{fig:CrackTipLeLd}b. Consequently, the stress state within this crack tip zone is elastic in nature. This finding is supported by a plot of tensile stress $\sigma_{yy}$ as a function of $r$ directly ahead of the crack tip ($y=0$), see Fig. \ref{fig:CrackTipLeLd}a. The stress component $\sigma_{yy}$ scales as $r^{-1/2}$ for sufficiently small $r$. Likewise, the elastic strain component $\varepsilon_{yy}^e$ scales as $r^{-1/2}$ for $r / \ell < 1$; from Hooke's law and Fig. \ref{fig:CrackTipLeLd} it is clear that the elastic strain dominates the plastic strain $\varepsilon_{yy}^e >> \varepsilon_{yy}^p$. Beyond the plastic zone ($r/R_p>1$) the stress state again converges to the elastic $K$-field and $\sigma_{yy}$ scales as $r^{-1/2}$. Thus, both an outer and an inner $K$ field exist. The plastic strain distribution $\varepsilon^p_{yy} (r)$ is relatively insensitive to the choice of values of $L_E$ and $L_D$ in Fig. \ref{fig:CrackTipLeLd}b. In all three cases, the plastic strain is almost constant over $0 < r < \ell$. In their recent asymptotic analysis, \citet{EJMaS2019} find that the leading order terms of the plastic strain $\varepsilon_{yy}^p$ along $\theta=0^\circ$ are
\begin{equation}
\varepsilon^p_{yy} = A + B r^{3/2}
\end{equation}
\noindent for the case $L_D \neq 0$, $L_E=0$, where $(A,B)$ are functions of $R_p$. In the present finite element study, it is also found that the plastic strain is finite at the crack tip when energetic higher order terms are present. We note in passing that the plastic strain is not sufficiently singular to contribute to the $J$-integral as the crack tip is approached. Instead, the $J$-integral is determined solely by the elastic strain state near the crack tip.\\ 

\begin{figure}[H]
  \makebox[\textwidth][c]{\includegraphics[width=0.77\textwidth]{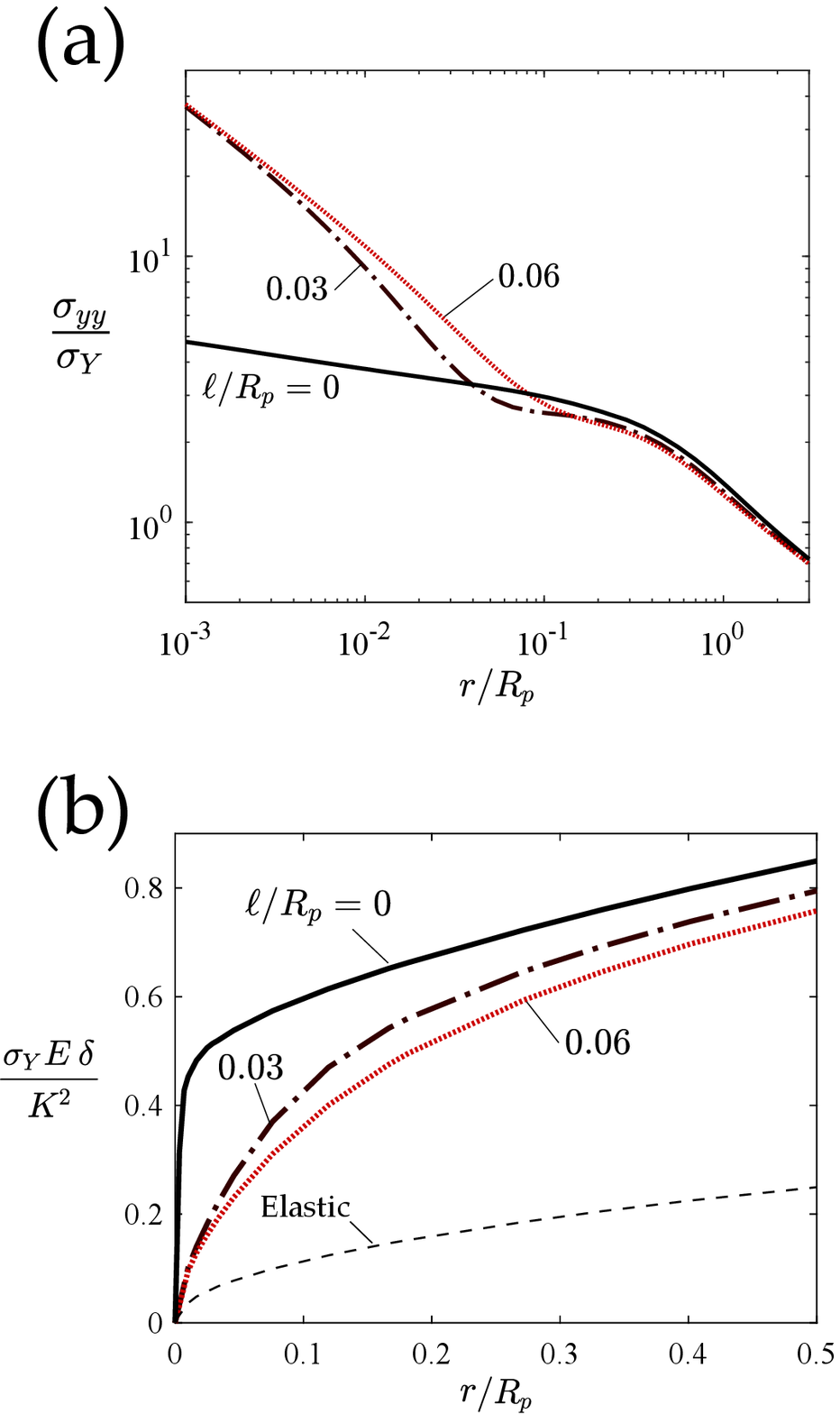}}%
  \caption{Response of a stationary crack for different length parameters $L_D=L_E=\ell$, (a) normalized tensile stress distribution ahead of the crack ($\theta=0^\circ$), and (b) crack tip opening profile ($\theta=180^\circ$). Material properties: $\sigma_Y/E=0.003$, $\nu=0.3$ and $N=0.1$. Small scale yielding conditions.}
  \label{fig:Stationary}
\end{figure}

The relative insensitivity of the stationary crack response in Fig. \ref{fig:CrackTipLeLd} to the ratio $L_E/L_D$ leads us to focus on a single reference length scale $L_E=L_D=\ell$. The tensile stress $\sigma_{yy}$ directly ahead of the crack tip is shown in Fig. \ref{fig:Stationary}a for selected values of $\ell/R_p$, with $\ell/R_p=0$ corresponding to the conventional plasticity limit. In all cases, except for $\ell/R_p=0$, the asymptotic stress state is elastic in nature, with the tensile stress exhibiting an $r^{-1/2}$ singularity. Now place a cohesive zone at the crack tip; then, a cohesive zone strength $\hat{\sigma}$ on the order of $4 \sigma_Y$ is sufficient to prevent crack advance in the conventional solid but not in the strain gradient case. In broad terms, the presence of strain gradients elevate stress and diminish the degree of plastic straining near the crack tip. To illustrate this, the crack opening profile for the strain gradient solid is compared to that of the conventional elasto-plastic solid ($\ell/R_p=0$) and to that of an elastic solid in Fig. \ref{fig:Stationary}b. The opening profile in the strain gradient plasticity solid ($\ell > 0$) is close to the elastic case as $r \to 0$, and is close to the conventional elasto-plastic solid as $r/R_p \to 1$.\\

The above results for the asymptotic crack tip fields are in marked contrast to those obtained by \citet{Chen1999}. They considered the asymptotic crack tip singular field for a mode I crack in a rigid power-law hardening strain gradient solid, as introduced by \citet{Fleck1997}. \citet{Chen1999} neglected elastic strains by assuming, a priori, that the crack tip plastic strain field dominates the elastic strains. They find that the crack tip plastic strain field scales as $r^{N/(N+1)}$ in order for the strain energy density to scale as $r^{-1}$ as the crack tip is approached (thereby giving a finite value of the $J$-integral at the crack tip). Consequently, the plastic strain vanishes as $r \to 0$. The asymptotic analysis of \citet{Chen1999} carries over directly to our case \emph{if we assume that elastic strains are negligible in comparison with plastic strains at the crack tip}. But in so doing, we find that the plastic strain vanishes at the crack tip and consequently the elastic strain vanishes at the crack tip also. This result is unphysical: the crack tip is sharp and will give rise to a strain concentration. We conclude that the elastic strains must dominate the plastic strains as the crack tip is approached. Consequently, an elastic $K$-field exists at the crack tip, such that the elastic strains and Cauchy stresses scale as $r^{-1/2}$. The finite element results fully support this finding, and reveal that the crack tip plastic strain is finite.\\

\citet{Chen1999} argued that the asymptotic field is not a physical representation over a small region (a small fraction of $\ell$) from the crack tip on the basis that the traction is negative in that zone.  We draw an alternative conclusion: within a zone of order $\ell$, the crack tip field is elastic in nature. The asymptotic field of \citet{Chen1999} has no zone of validity as it neglects elastic straining.

\section{Analysis of a growing crack}
\label{Sec:Results}

\begin{figure}[H]
  \makebox[\textwidth][c]{\includegraphics[width=0.85\textwidth]{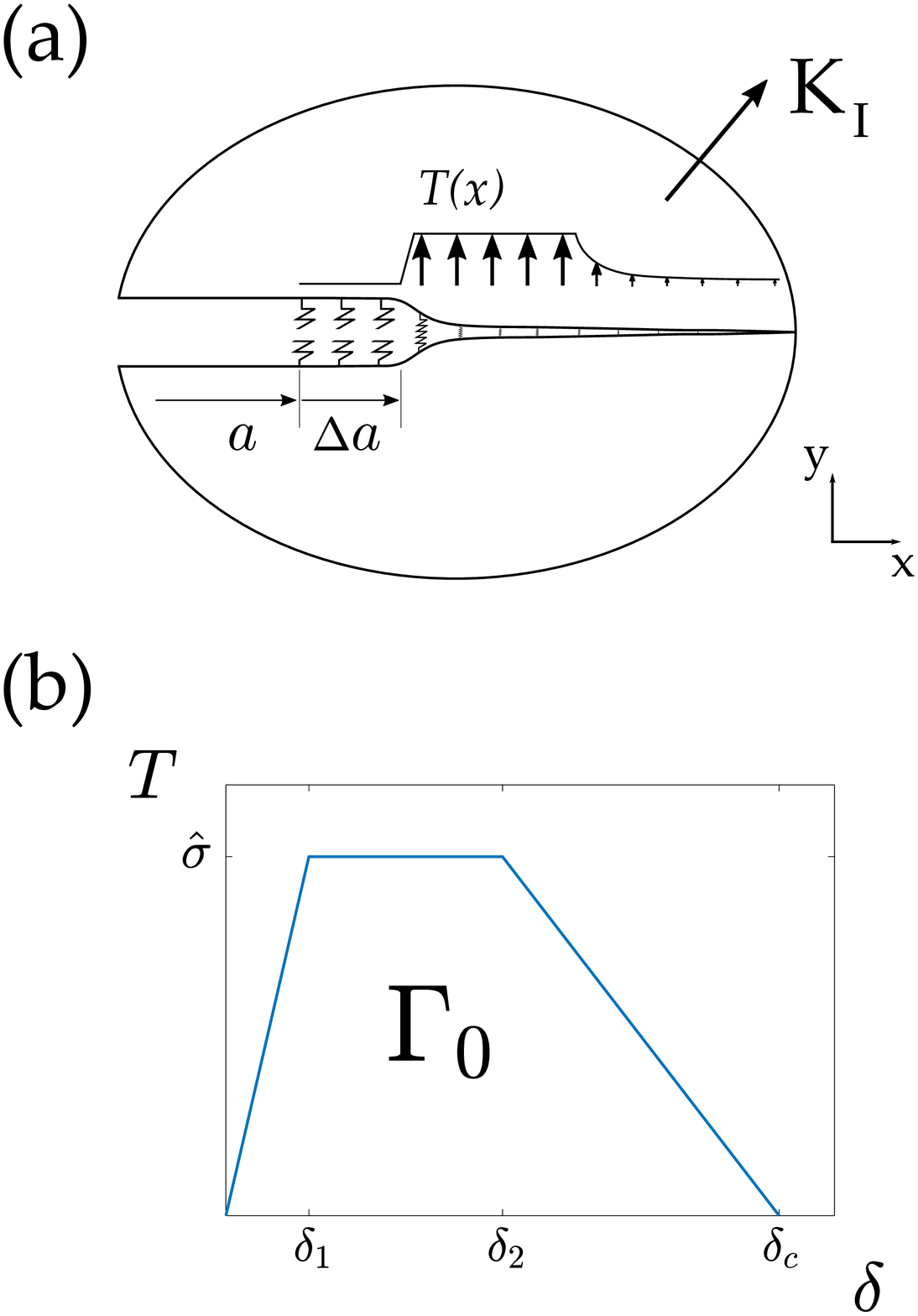}}%
  \caption{Cohesive zone description of fracture, (a) schematic representation, and (b) constitutive traction-separation relation.}
  \label{fig:CZM}
\end{figure}

In the current study, we investigate crack growth from either a short or a long crack by making use of strain gradient plasticity theory. In the long crack case, R-curves are obtained and the present study thereby extends the results of \citet{Tvergaard1992} by incorporating the role of plastic strain gradients. Failure by cleavage, by void growth or by other mechanisms is idealised by an assumed traction $T$ versus separation $\delta$ law along a cohesive strip directly ahead of the crack tip, see Fig. \ref{fig:CZM}a. Following \citet{Tvergaard1992}, a trapezoidal shape is assumed for the $T(\delta)$ relation, as characterised by three salient values of opening ($\delta_1$, $\delta_2$, $\delta_c$) and a strength $\hat{\sigma}$ (see Fig. \ref{fig:CZM}b). We hold fixed the ratios $\delta_1/\delta_c=0.15$ and $\delta_2/\delta_c=0.5$, and thereby treat $(\delta_c, \hat{\sigma})$ as the two primary parameters that define the cohesive zone law. The work of fracture $\Gamma_0$ is the area under the $T (\delta)$ curve, as given by
\begin{equation}\label{Eq:CoheLaw}
\Gamma_0=\frac{1}{2} \hat{\sigma} \left( \delta_c + \delta_2 - \delta_1 \right)
\end{equation}

It follows directly from the surface work terms on the right hand side of (\ref{eq:PVW1Gud}) that, in general, a cohesive zone can support both tractions $T_i$ and higher order tractions $t_{ij}$. We assume that the tensile traction $T$ on the cohesive zone depends only upon the crack opening displacement $\delta$. Further, we assume that the higher order traction $t_{ij}$ vanishes on the surface of the cohesive zone; this is a natural boundary condition in the finite element formulation. The use of a cohesive zone model embedded within an elasto-plastic solid gives insight into both ductile fracture and cleavage by suitable choices of the cohesive zone parameters $\hat{\sigma}$ and $\Gamma_0$.\\

\begin{figure}[H]
  \makebox[\textwidth][c]{\includegraphics[width=0.9\textwidth]{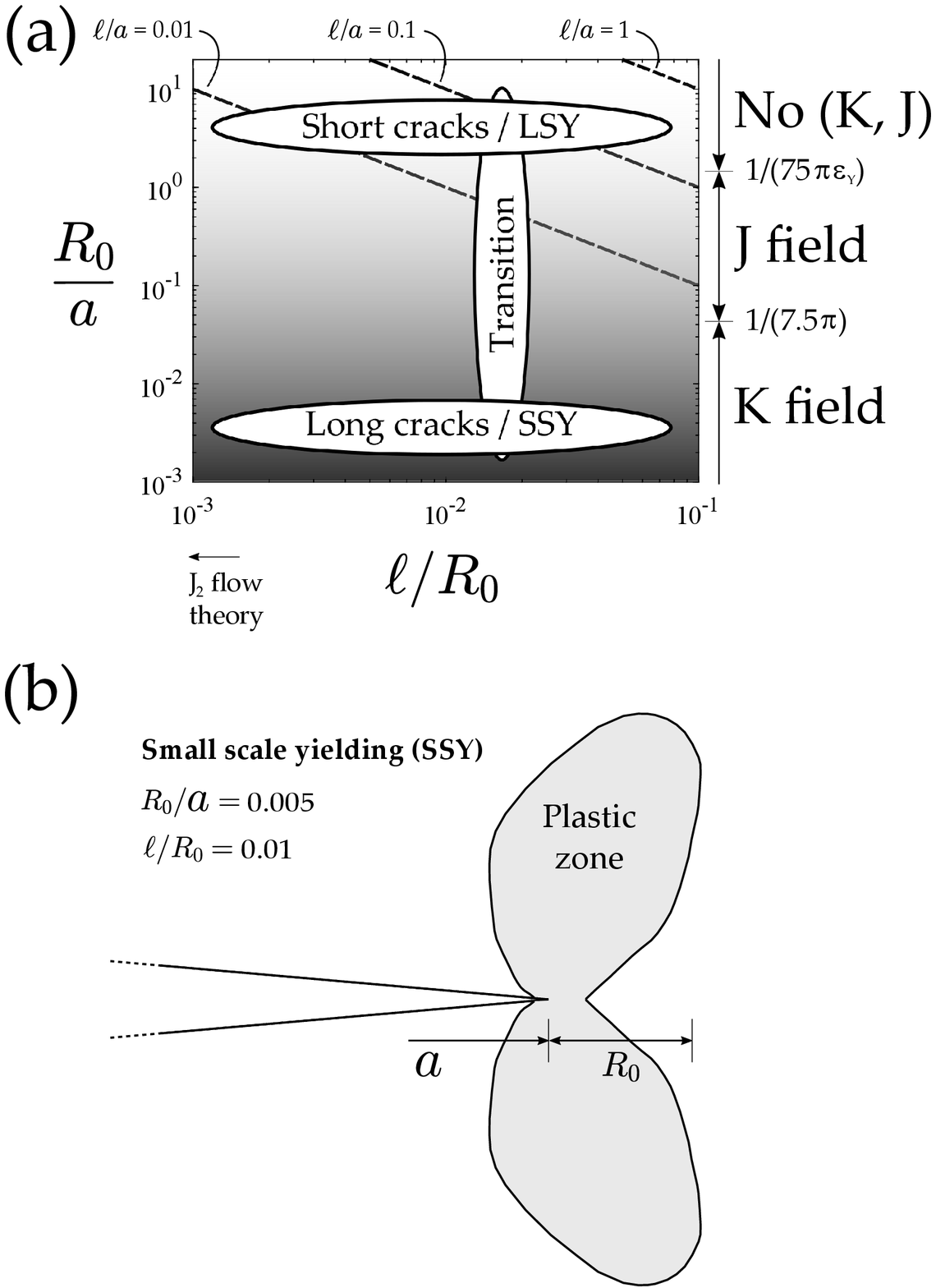}}%
  \caption{Schematic diagram of the regimes and competing length scales involved in the fracture process of metals. Material properties: $\sigma_Y/E=0.003$.}
  \label{fig:LengthScales}
\end{figure}

We proceed to evaluate the influence of crack length, material length scale $\ell$ of the strain gradient solid, and a representative fracture process zone size\footnote{$R_0$ corresponds to the plastic zone size in a conventional solid at the onset of crack growth.}
\begin{equation}\label{Eq:R0}
R_0 = \frac{1}{3 \pi (1 - \nu^2)} \frac{E \Gamma_0}{\sigma_Y^2}
\end{equation}

\noindent on the fracture response. Regimes of behaviour are sketched in non-dimensional space $(\ell/R_0, R_0/a)$ in Fig. \ref{fig:LengthScales}a. Our analysis spans the regimes of small scale yielding (for which an outer $K$-field exists), $J$-controlled fracture and large scale plasticity. A representative crack tip plastic zone, computed at crack initiation, is shown in Fig. \ref{fig:LengthScales}b for the case of small scale yielding. The plastic zone size is defined by the contour along which the von Mises effective stress equals the initial yield stress. Crack growth resistance is assessed for three distinct regimes in $(\ell/R_0, R_0/a)$ space, as shown by the ellipses in Fig. \ref{fig:LengthScales}a. Section \ref{Sec:Rcurves} deals with the fracture response of a strain gradient plasticity solid with a long crack while the mechanics of short flaws and the influence of crack length on the fracture response are addressed in Section \ref{Sec:ShortCrack}.

\subsection{Boundary value problem}
\label{Sec:Plate}

\begin{figure}[H]
\centering
\includegraphics[scale=1.2]{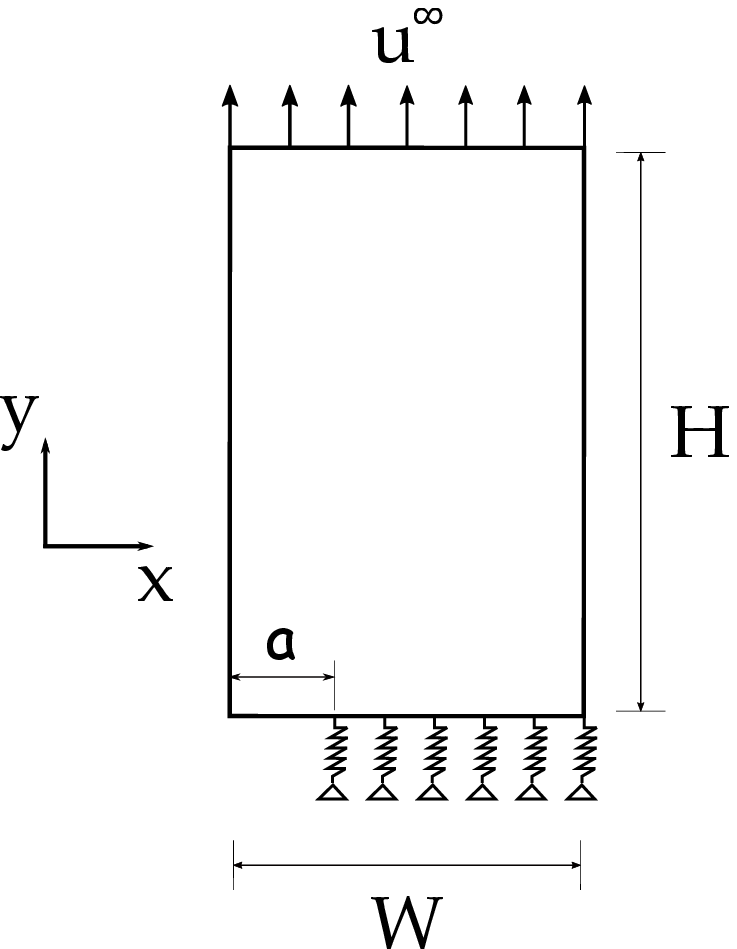}
\caption{Configuration of the edge cracked plate employed to model crack growth in the presence of short and long cracks.}
\label{fig:Plate}
\end{figure}

We investigate crack initiation and subsequent growth in an edge-cracked plate loaded in uniaxial tension under plane strain conditions, see Fig. \ref{fig:Plate}. The same geometry is used for the study of long and short pre-cracks. The specimen has a height-to-width ratio of $H/W=4$ and an initial crack length of $a/W=0.1$. The cohesive zone model outlined above is employed to model crack initiation and growth. Following \citet{Wei1997}, micro-free boundary conditions $t_{ij}=0$ are adopted on the symmetry plane. Cohesive elements with 6 nodes and 12 integration points are implemented by means of a user element (UEL) subroutine, as described elsewhere \citep{EFM2017}. The finite element mesh is refined ahead of the initial crack tip to ensure that the element size is able to resolve the fracture process zone. Specifically, the model consists of approximately $10^6$ degrees of freedom and the characteristic element length equals $R_0/100$. Post-processing of the results is performed with Abaqus2Matlab \citep{AES2017}. \\

It is widely appreciated that elastic snap-back instabilities can arise when cohesive elements experience stiffness degradation, complicating the modelling of the post-instability behaviour. The simultaneous reduction of the remote load and the prescribed displacement inevitably triggers convergence problems in quasi-static finite element computations. A numerical strategy to overcome these instabilities lies in prescribing a quantity that increases monotonically throughout the loading history while making the remote load an output of the model (\citealp{Tvergaard1976}; \citealp{Segurado2004}). In the present study, a control algorithm is used to prescribe the crack tip opening and obtain the displacement at the remote boundary by ensuring global force equilibrium. Details are given in the Supplementary Material.

\subsection{Small scale yielding response}
\label{Sec:Rcurves}
The $K$ calibration for the specimen geometry of Fig. \ref{fig:Plate} was determined as follows. Consider the elastic solid, absent a cohesive zone and apply a uniform remote displacement $u_y=u^\infty$ on the top edge, with $T_x\equiv 0$. A linear elastic finite element calculation reveals that the average traction $\bar{T}$ on the top edge is $\bar{T} \approx E' u^\infty/H$ where $E'=E/(1-\nu^2)$, assuming plane strain conditions. A contour integral evaluation of the stress intensity factor $K$ at the crack tip gives $K=1.15 \bar{T} \sqrt{\pi a}$. Thus, for the small scale yielding case of limited crack tip plasticity, the remote $K$ value for the geometry of Fig. \ref{fig:Plate} is given by,
\begin{equation}\label{Eq:Kremote}
K=\frac{1.15 E u^\infty \sqrt{\pi a}}{(1-\nu^2) H}
\end{equation}

Small scale yielding prevails when,
\begin{equation}\label{Eq:CrackSizeASTM}
a > 2.5 \frac{K^2}{\sigma_Y^2}
\end{equation}

\noindent in accordance with ASTM E1820. This places an upper limit on the value of $u^\infty/H$ for small scale yielding; rearrangement of (\ref{Eq:Kremote}) and (\ref{Eq:CrackSizeASTM}) implies,
\begin{equation}\label{Eq:Uinfty}
\frac{u^\infty}{H} < \frac{\sigma_Y (1-\nu^2)}{1.8 \sqrt{\pi} E }
\end{equation}

\noindent This condition was satisfied in the following determination of the R-curve under small scale yielding conditions. Consider a long crack subjected to a remote load $K$. Crack initiation occurs within the cohesive zone at a value of $K$ equal to
\begin{equation}\label{Eq:K0}
K_0 = \left( \frac{E \Gamma_0}{1 - \nu^2 } \right)^{1/2}
\end{equation}

\begin{figure}[H]
  \makebox[\textwidth][c]{\includegraphics[width=0.75\textwidth]{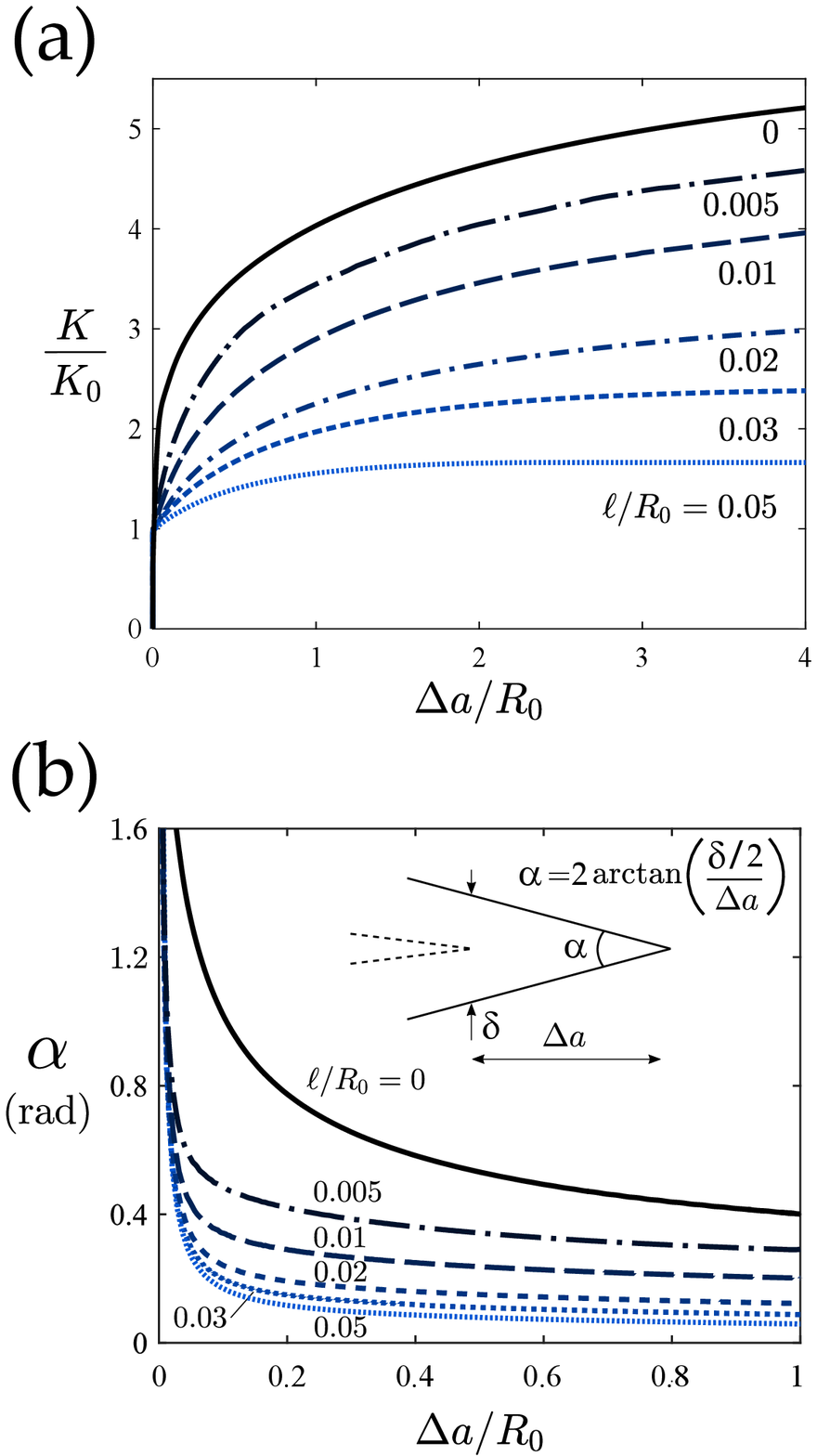}}%
  \caption{Crack growth resistance for different length parameters $L_D=L_E=\ell$, (a) R-curves, and (b) crack opening angle. Long crack $a/R_0=125$. Material properties: $\delta_1/\delta_c=0.15$, $\delta_2/\delta_c=0.5$, $\hat{\sigma}/\sigma_Y=3.8$, $\sigma_Y/E=0.003$, $\nu=0.3$ and $N=0.1$.}
  \label{fig:Rcurves}
\end{figure}

Dimensional analysis implies that the crack growth resistance for a long crack depends on the following dimensionless groups
\begin{equation}
\frac{K}{K_0}=F \left( \frac{\Delta a}{R_0}, \, \frac{\hat{\sigma}}{\sigma_Y}, \frac{\ell}{R_0}; \, N, \, \frac{\sigma_Y}{E}, \, \nu \right)
\end{equation}

\noindent where $(N, \, \sigma_Y/E, \, \nu)$ are held fixed in the present study, along with the values of $\delta_1/\delta_c$ and $\delta_2 /\delta_c$ in (\ref{Eq:CoheLaw}). The computed crack growth resistance curves for $\hat{\sigma}/\sigma_Y=3.8$ and for selected values of the constitutive length scales $L_E=L_D=\ell$, relative to $R_0$, are shown in Fig. \ref{fig:Rcurves}a. The influence of plastic strain gradients in lowering the fracture resistance is evident: the steepness of the R-curve and the steady state value $K_{SS}/K_0$ diminish with increasing $\ell/R_0$. \citet{Seiler2016} considered the initial stage of the R-curve for a visco-plastic solid whereby the viscoplastic strain rate $\dot{\varepsilon}^{VP}$ scales as $\sigma^m$ where $1<m<\infty$. They showed that the sensitivity of the R-curve to the material length scale $\ell$ increases with increasing $m$. In the present study we consider the rate independent limit, $m \to \infty$, and a high sensitivity of the R-curve to length $\ell$ is, indeed, observed.\\

The crack tip opening angle has been used as a criterion for crack growth resistance in metallic alloys \citep{Kanninen1985}. The dependence of the crack tip opening angle upon crack extension is shown in Fig. \ref{fig:Rcurves}b. Here, the crack opening angle $\alpha$, as defined in the inset of Fig. \ref{fig:Rcurves}b, is almost independent of $\Delta a$ after an initial transient phase. The steady state value of $\alpha$ decreases with increasing $\ell/R_0$, consistent with the crack opening profile for a stationary crack, as shown in Fig. \ref{fig:Stationary}b. It is clear that the crack tip opening angle is sensitive to strain gradient effects. In turn, this is due to the sensitivity of the plastic strain field to strain gradient effects. This is now explored in detail.\\

\begin{figure}[H]
  \makebox[\textwidth][c]{\includegraphics[width=0.9\textwidth]{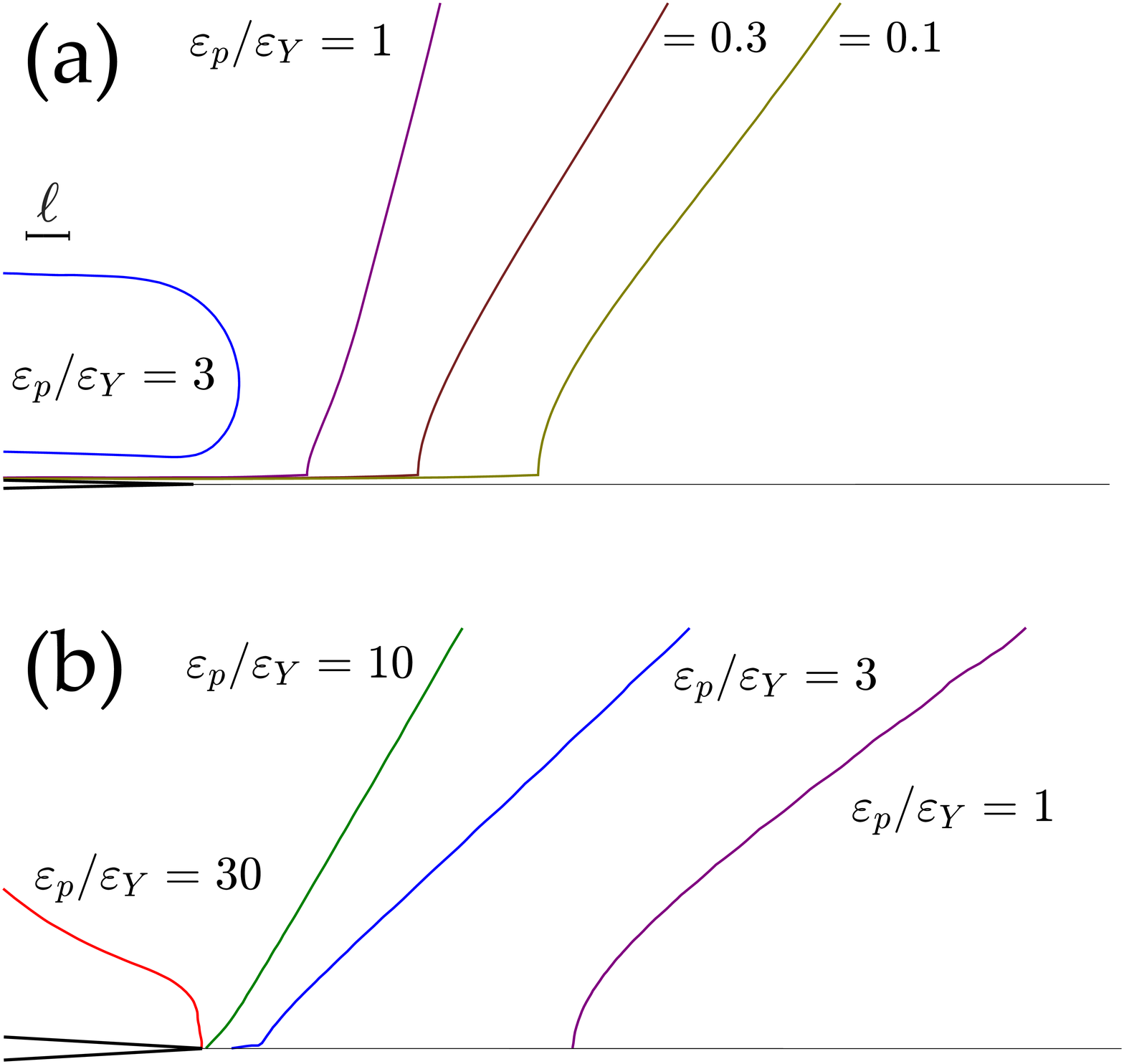}}%
  \caption{Effective plastic strain contours ahead of a propagating crack at steady state, (a) strain gradient plasticity, with $L_D=L_E=\ell=0.05R_0$, and (b) conventional plasticity. Long crack $a/R_0=125$. Material properties: $\delta_1/\delta_c=0.15$, $\delta_2/\delta_c=0.5$, $\hat{\sigma}/\sigma_Y=3.8$, $\sigma_Y/E=0.003$, $\nu=0.3$ and $N=0.1$.}
  \label{fig:ElasticCore}
\end{figure}

The plastic field surrounding the tip of a crack propagating at steady state is examined in Fig. \ref{fig:ElasticCore}. A von Mises measure of plastic strain is defined as,
\begin{equation}
\varepsilon_p = \left( \frac{2}{3} \varepsilon_{ij}^p \varepsilon_{ij}^p \right)^{1/2}
\end{equation}
\noindent and its contours are plotted in Fig. \ref{fig:ElasticCore} for strain gradient plasticity, with $\ell/R_0=0.05$, and also for the conventional plasticity case $\ell=0$. For the choice $\ell/R_0=0.05$, plastic strains attain a plateau value of $\varepsilon_p/\varepsilon_Y=3$ at a distance on the order of $\ell$ from the crack tip. Furthermore, the maximum level of plastic strain is not attained at the crack tip, a feature which also observed in discrete dislocation plasticity \citep{Chakravarthy2010}. This contrasts with the conventional plasticity case, see Fig. \ref{fig:ElasticCore}b. In addition, plastic strains are approximately one order of magnitude larger than for the strain gradient plasticity case.\\

\begin{figure}[H]
\centering
\includegraphics[scale=1.1]{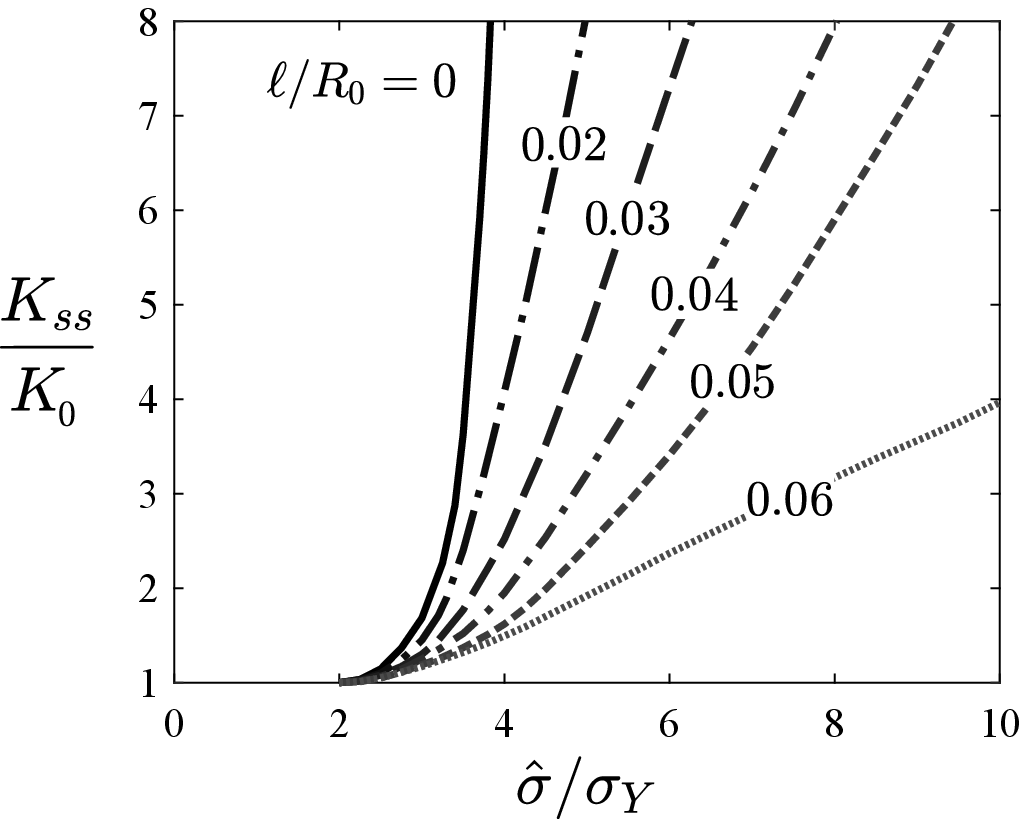}
\caption{Steady state toughness as a function of $\hat{\sigma}/\sigma_Y$ for different length parameters $L_D=L_E=\ell$. Long crack $a/R_0=125$. Material properties: $\delta_1/\delta_c=0.15$, $\delta_2/\delta_c=0.5$, $\sigma_Y/E=0.003$, $\nu=0.3$ and $N=0.1$.}
\label{fig:SScurves}
\end{figure}

The dependence of $K_{SS}/K_0$ upon $\hat{\sigma}/\sigma_Y$ is given in Fig. \ref{fig:SScurves} for selected values of $\ell/R_0$. There is a qualitative change when $\ell/R_0$ is increased from zero to a finite value. For $\ell/R_0=0$, continued crack advance (at $K=K_{SS}$) is precluded for $\hat{\sigma}/\sigma_Y>4$; the level of crack tip stress is unable to overcome the cohesive strength when $\hat{\sigma}/\sigma_Y \geq 4$. In contrast, when strain gradients are taken into account, the crack tip stresses can attain any value of cohesive strength, and $K_{SS}/K_0$ increases monotonically with increasing $\hat{\sigma}/\sigma_Y$. However, the degree of elevation of the R-curve, $K_{SS}/K_0$, decreases with increasing $\ell/R_0$ for any given $\hat{\sigma}/\sigma_Y$; this is consistent with the results shown in Fig. \ref{fig:Rcurves}a for the choice $\hat{\sigma}/\sigma_Y=3.8$. Recall that the choice of $\hat{\sigma}/\sigma_Y \approx 10$ is representative of the mechanism of quasi-cleavage in metallic alloys: the crack tip advances by cleavage, but surrounded by a plastic zone. The predictions of Fig. \ref{fig:SScurves} show that a shallow R-curve can exist for such a case: $K_{SS}/K_0$ equals 4 for $\hat{\sigma}/\sigma_Y=10$ and $\ell/R_0=0.06$. The qualitative response is similar to that obtained by \citet{Wei1997} for the case of \citet{Fleck1997} strain gradient theory. However, significant quantitative differences arise. If we consider a cohesive strength of $\hat{\sigma}/\sigma_Y  \approx 10$ in both studies, then a value of $K_{SS}/K_0$ on the order of 4 is achieved for $\ell/R_0$ an order of magnitude smaller than that found by \citet{Wei1997}. \\

\begin{figure}[H]
\centering
\includegraphics[scale=1.1]{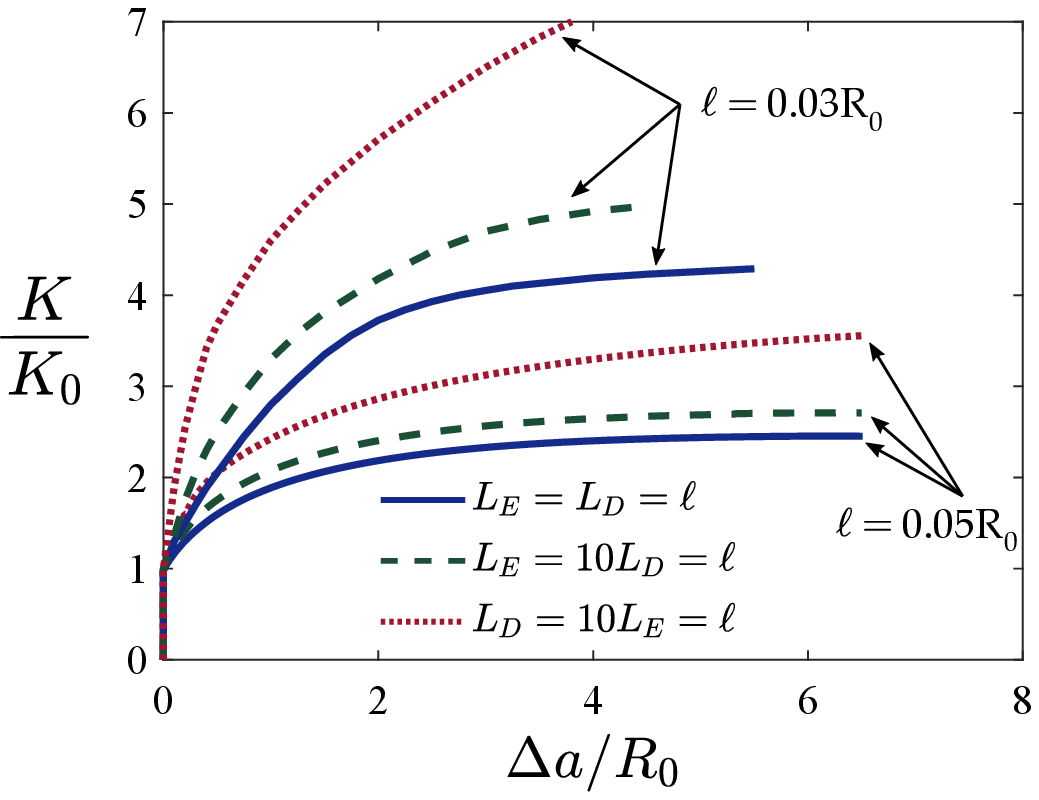}
\caption{Crack growth resistance curves for different combinations of length scale parameters: $L_E=L_D=\ell$, $L_E=10L_D=\ell$, and $L_D=10L_E=\ell$. Long crack $a/R_0=125$. Material properties: $N=0.1$, $\sigma_Y/E=0.003$, $\hat{\sigma}/\sigma_Y=5$ and $\nu=0.3$.}
\label{fig:LevsLd}
\end{figure}

Finally, we investigate the relative influence of energetic and dissipative gradient contributions to the R-curve. Crack growth resistance curves are shown in Fig. \ref{fig:LevsLd}a for three cases: (i) $L_E=10L_D=\ell$, (ii) $L_D=10L_E=\ell$, and (iii) $L_D=L_E=\ell$ (i.e., the reference case). All of the R-curves are for $\hat{\sigma}/\sigma_Y=5$, and results are given for the two choices $\ell/R_0=0.03$ or $\ell/R_0=0.05$. The R-curve is steepest for $\ell/R_0=0.03$ and $L_D=10L_E=\ell$, for which dissipative hardening dominates. Combined energetic and dissipative hardening with $L_E=L_D=\ell$ emphasizes the role of strain gradients and leads to a less steep R-curve; the choice $L_E=10L_D=\ell$ (energetic hardening dominant) is the intermediate case. Consistent with the results shown in Fig. \ref{fig:SScurves}, for which $L_E=L_D=\ell$, the R-curve is less steep and $K_{SS}/K_0$ drops with increasing $\ell/R_0$ for all 3 choices of $L_D/L_E$.\\

\subsection{Short crack limit}
\label{Sec:ShortCrack}

We now turn our attention to crack advance from a short pre-crack, for which $a/R_0<1$, recall Fig. \ref{fig:LengthScales}a. Such cracks commonly arise at grain boundaries, at cracked carbide particles or as machining damage in structural alloys. The cracks are sufficiently short for no $K$-field (or $J$-field) to exist and are accompanied by plastic collapse at the structural level. Thus, failure occurs at a stress level somewhat above the yield strength, and the question of interest becomes: what is the dependence of macroscopic failure strain (below the necking strain) on $a/R_0$ and $\ell/R_0$?\\

\begin{figure}[H]
  \makebox[\textwidth][c]{\includegraphics[width=0.85\textwidth]{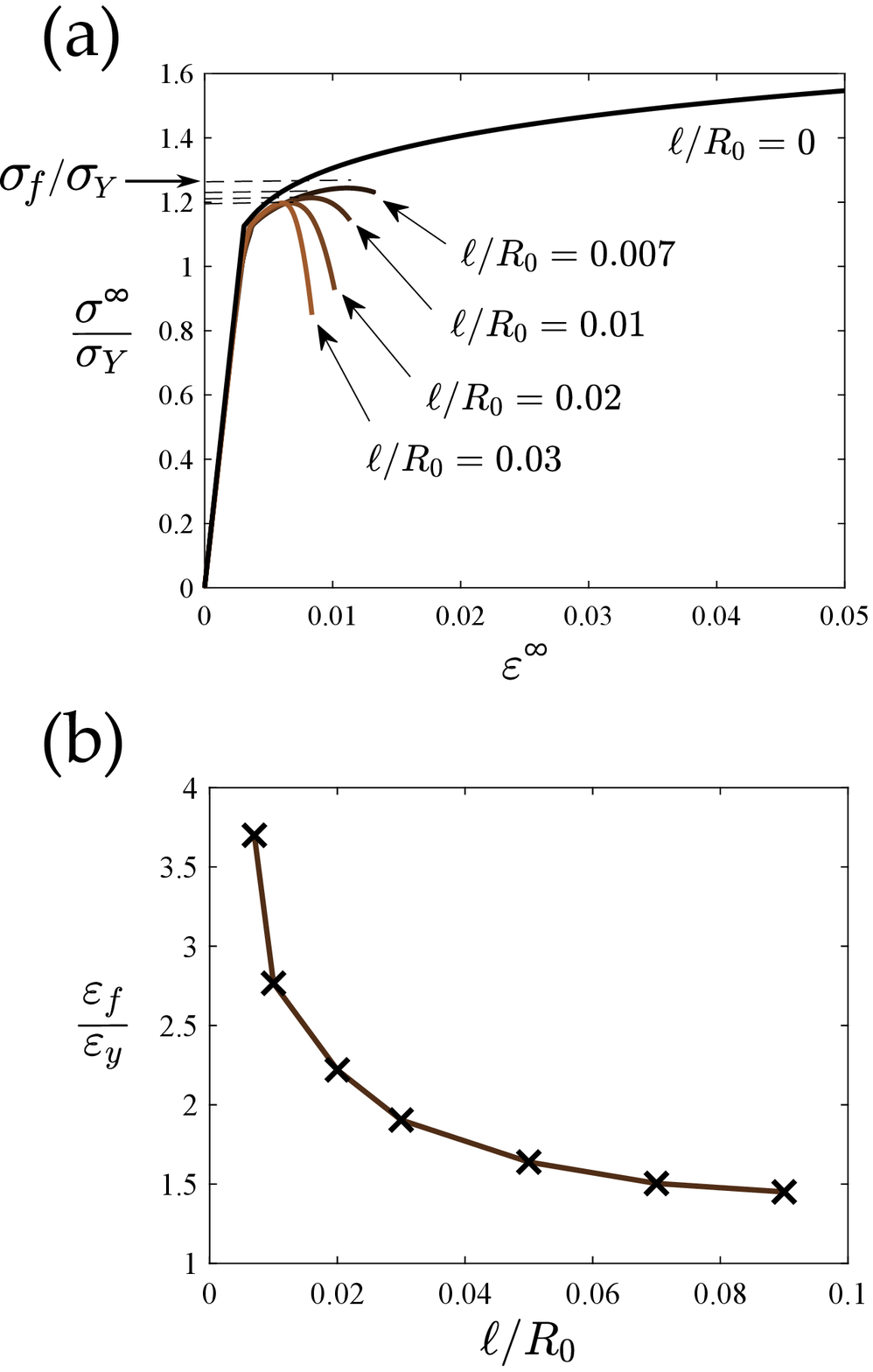}}%
  \caption{Fracture response in short cracks ($a_0/R_0=0.38$): (a) remote stress versus nominal strain, and (b) failure strain versus length scale parameter. Material properties: $\hat{\sigma}/\sigma_Y=5$, $\delta_1/\delta_c=0.15$, $\delta_2/\delta_c=0.5$, $\sigma_Y/E=0.003$, $\nu=0.3$ and $N=0.1$.}
  \label{fig:SrvsE}
\end{figure}

First consider the case of a short crack of length $a/R_0=0.38$. Dimensional analysis implies,
\begin{equation}
\frac{\sigma^\infty}{\sigma_Y}=F \left( \frac{\varepsilon^\infty}{\varepsilon_y}, \frac{\ell}{R_0}, \, \frac{a}{R_0}; \, \frac{\hat{\sigma}}{\sigma_Y}, \, \frac{\sigma_Y}{E}, \, \nu, \, N \right)
\end{equation}

\noindent where $\sigma^\infty$ is the macroscopic remote stress on a tensile specimen (recall Fig. \ref{fig:Plate}) and $\varepsilon^\infty$ is the work-conjugate remote tensile strain. A series of finite element simulations have been performed for $\hat{\sigma}/\sigma_Y=5$ and $N=0.1$, for illustrative purposes. The $\sigma^\infty$ versus $\varepsilon^\infty$ response is given in Fig. \ref{fig:SrvsE}a for selected values of $\ell/R_0$ in the range 0 to 0.03. The tensile response is very sensitive to the choice of $\ell/R_0$, as follows. For $\ell/R_0=0$, the tensile response is almost identical to the material stress versus strain curve, and no failure is predicted. In contrast, the failure strain drops to about 1\% when plastic strain gradients are accounted for. This is emphasized by the plot of failure strain $\varepsilon_f$ versus $\ell/R_0$ in Fig. \ref{fig:SrvsE}b: $\varepsilon_f$ drops steeply from $\varepsilon_f / \varepsilon_y=3.7$ at $\ell/R_0=0.007$ to $\varepsilon_f / \varepsilon_y=1.45$ at $\ell/R_0=0.09$. Thus, strain gradient plasticity theory, along with a cohesive zone model, gives mechanistic insight into the drop in ductility when the fracture length scale $R_0$ drops (e.g., due to embrittlement) in relation to the plasticity length scale $\ell$.\\

\begin{figure}[H]
  \makebox[\textwidth][c]{\includegraphics[width=0.85\textwidth]{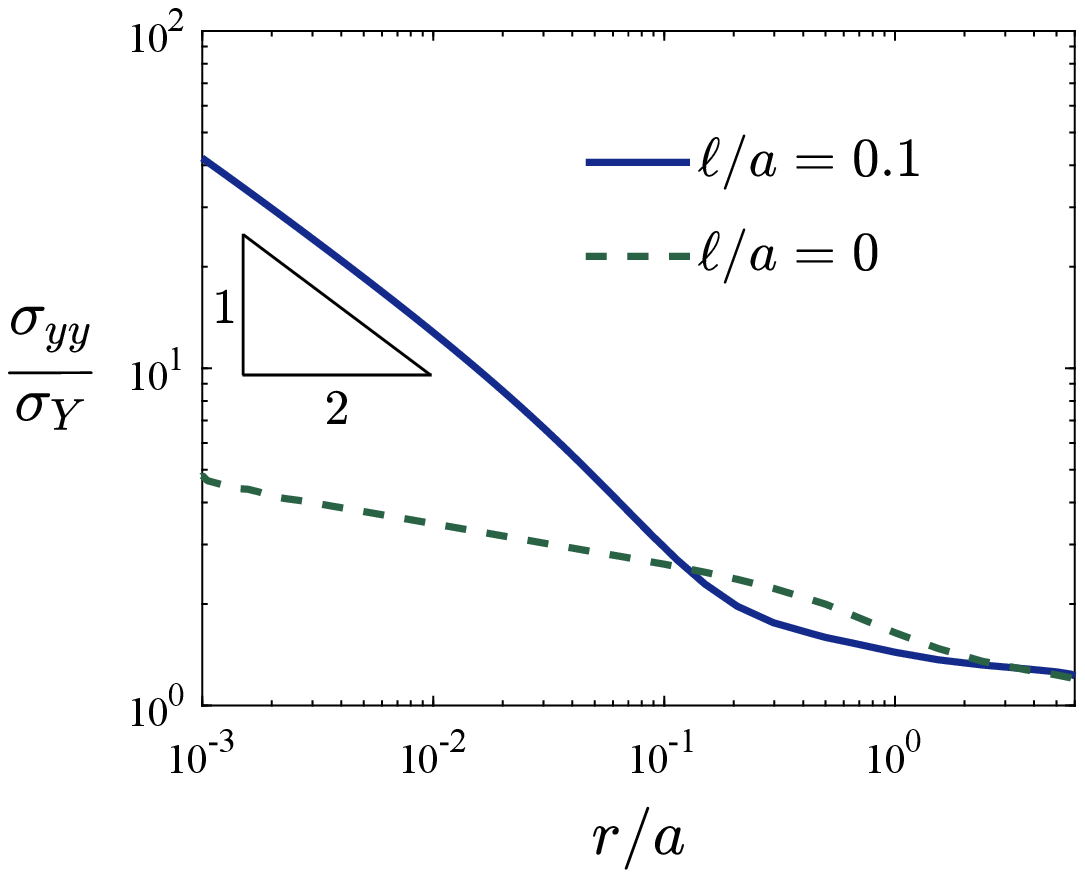}}%
  \caption{Tensile stresses ahead of a stationary short crack ($\theta=0^\circ$) for $\ell/a=0.1$ and the conventional case, $\ell/a=0$. Remote tensile strain $\varepsilon^\infty=0.005$. Material properties: $\sigma_Y/E=0.003$, $N=0.1$, and $\nu=0.3$.}
  \label{fig:ShortSta}
\end{figure}

In order to interrogate the source of the dramatic drop in ductility with increasing $\ell$ we examine the stress field ahead of a stationary short crack, absent the cohesive zone. The tensile stress $\sigma_{yy}$ is plotted as a function of $r/a$ in Fig. \ref{fig:ShortSta} for a fixed value of $\varepsilon^\infty=0.005$ such that material remote from the crack tip has fully yielded. Results are shown for $\ell/a=0, 0.1$ and it is clear that the asymptotic stress field is similar to that for the long crack, as shown in Fig. \ref{fig:CrackTipLeLd}. For the strain gradient solid, an elastic zone of extent on the order of $\ell$ exists at the crack tip. A crack tip $K$-field is evident, as for the long crack case, and it is this feature that results in the drop in ductility for the growing crack case of Fig. \ref{fig:SrvsE}.\\

It remains to explore the dependence of failure strength $\sigma_f/\sigma_Y$ upon crack length $a/R_0$. It is anticipated that, for sufficiently large $a/R_0$, small scale yielding applies and failure occurs at $K=K_{SS}$ for a long pre-crack, such that $\sigma_f \approx K_{SS}/\sqrt{\pi a}$. With diminishing crack length, $\sigma_f/\sigma_Y$ rises until, for sufficiently small pre-cracks ($a/R_0<25$) the $K$-field ceases to exist and a $J$-analysis is necessary for a fracture mechanics assessment. A further reduction in $a/R_0$ leads to the short crack regime, and the full trajectory of $a/R_0$ is labelled as \emph{transition} in Fig. \ref{fig:LengthScales}a. The above qualitative discussion is now made precise by a series of calculations for selected values of $a/R_0$.\\

\begin{figure}[H]
  \makebox[\textwidth][c]{\includegraphics[width=0.77\textwidth]{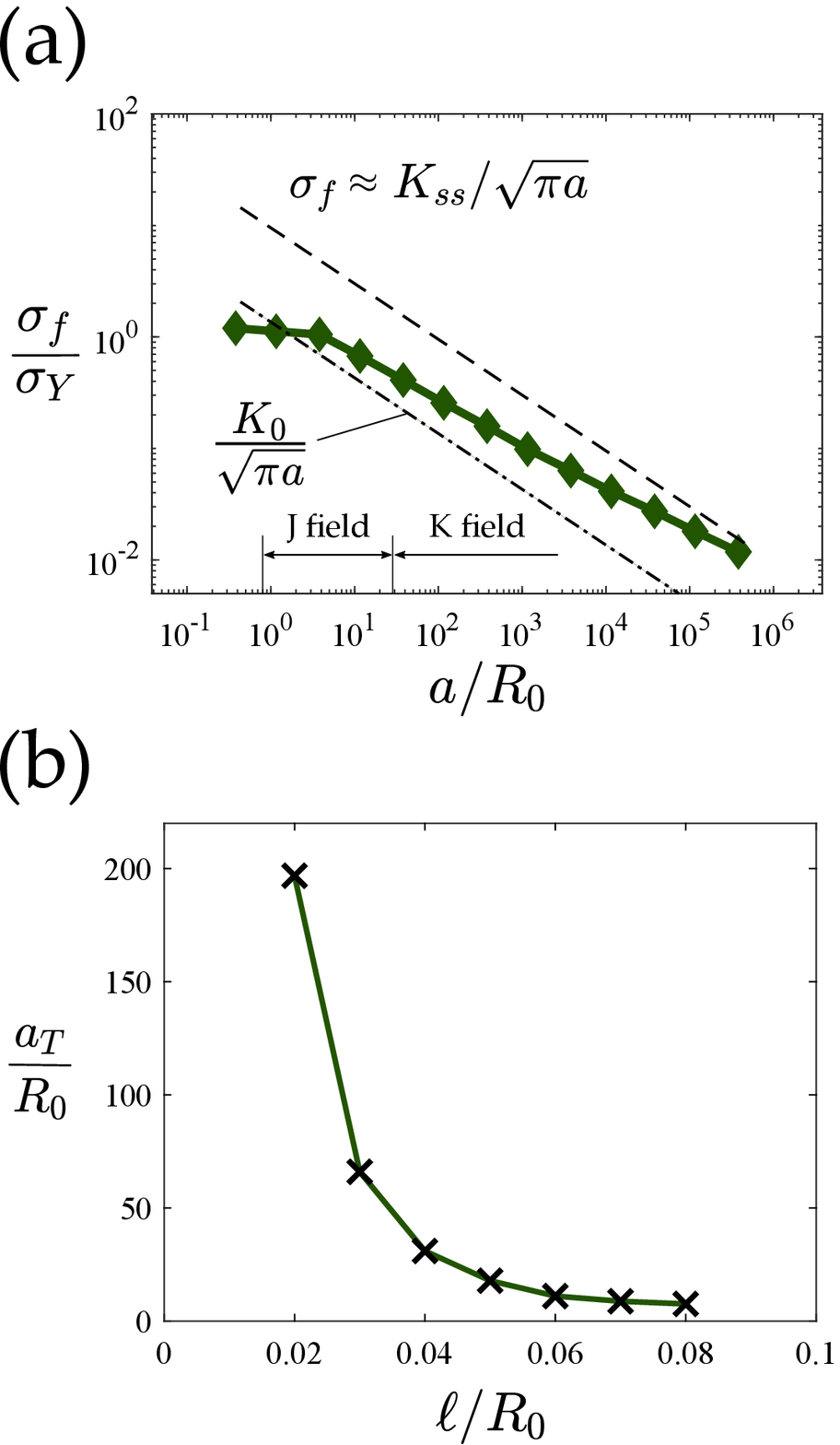}}%
  \caption{Influence of the crack length: (a) failure stress versus crack length for $\ell/R_0$, and (b) transition flaw sensitivity to $\ell/R_0$. Material properties: $\hat{\sigma}/\sigma_Y=5$, $\delta_1/\delta_c=0.15$, $\delta_2/\delta_c=0.5$, $\sigma_Y/E=0.003$, $\nu=0.3$ and $N=0.1$.}
  \label{fig:Transition}
\end{figure}

The predicted failure strength $\sigma_f/\sigma_Y$ is plotted as a function of $a/R_0$ in Fig. \ref{fig:Transition}a for the case of fixed $\ell/R_0=0.02$. As expected, $\sigma_f/\sigma_Y$ increases from the small scale yielding value to the plastic collapse value $\sigma_f/ \sigma_Y \approx 1$ with diminishing $a/R_0$. The regimes of validity of $K$ and $J$ are shown for completeness. A transition crack length can be identified by equating the fracture strength from plastic collapse theory $\sigma_f=\sigma_Y$ to the fracture strength from $K=K_{SS}$; such that $\sigma_Y \sqrt{\pi a_T}=K_{SS}$. Thus,
\begin{equation}\label{Eq:a_T}
a_T \equiv \frac{1}{\pi} \left( \frac{K_{SS}}{\sigma_Y} \right)^2
\end{equation}

The dependence of $a_T/R_0$ upon $\ell/R_0$ is plotted in Fig. \ref{fig:Transition}b upon making use of (\ref{Eq:a_T}). Upon recalling (\ref{Eq:R0}), the relation (\ref{Eq:a_T}) reduces to 
\begin{equation}
\frac{a_T}{R_0} = 3 \left( \frac{K_{SS}}{K_0} \right)^2
\end{equation}

Thus, the sensitivity of $a_T/R_0$ to $\ell/R_0$ arises directly from the dependence of $K_{SS}/K_0$ upon $\ell/R_0$.

\section{Conclusions}
\label{Sec:Concluding remarks}

The current study highlights the role of plastic strain gradients in influencing the R-curve for a long crack under small scale yielding and the tensile response in the presence of a short crack. An asymptotic analysis of the elastic-plastic stress state at the tip of a stationary crack in a strain gradient solid reveals that an elastic zone is present in the immediate vicinity of the crack tip. Consequently, the tensile stress immediately ahead of the crack tip displays an inverse square root singularity, in contrast to the HRR field of a conventional solid. This has immediate implications for a cohesive zone analysis of a growing crack: crack advance is predicted for cohesive strengths much greater than the yield strength. These predictions are consistent with observations of quasi-cleavage fracture with limited plasticity (\citealp{Elssner1994}; \citealp{Bagchi1996}; \citealp{Korn2002}).\\

Our study also reveals that the elastic crack tip singularity persists for the short crack case. Consequently, the tensile stress ahead of the short crack can far exceed the yield strength and overcome the cohesive strength of a cohesive zone placed at the crack tip. In turn, this leads to a significant drop in tensile ductility.
 
\section{Acknowledgments}
\label{Acknowledge of funding}

The authors would like to acknowledge the funding and technical support from BP (ICAM02ex) through the BP International Centre for Advanced Materials (BP-ICAM). E. Mart\'{\i}nez-Pa\~neda also acknowledges financial support from the Ministry of Economy and Competitiveness of Spain through grant MAT2014-58738-C3 and the People Programme (Marie Curie Actions) of the European Union's Seventh Framework Programme (FP7/2007-2013) under REA grant agreement n$^{\circ}$ 609405 (COFUNDPostdocDTU). C.F. Niordson acknowledges support from the Danish Council for Independent Research through the research project ``Advanced Damage Models with InTrinsic Size Effects" (Grant no: DFF-7017-00121). N.A. Fleck and V.S. Deshpande are grateful for additional financial support from the European Research Council in the form of an Advance Grant (MULTILAT, 669764).


\bibliographystyle{elsarticle-harv}
\bibliography{library}

\begin{thebibliography}{56}
\expandafter\ifx\csname natexlab\endcsname\relax\def\natexlab#1{#1}\fi
\expandafter\ifx\csname url\endcsname\relax
  \def\url#1{\texttt{#1}}\fi
\expandafter\ifx\csname urlprefix\endcsname\relax\def\urlprefix{URL }\fi

\bibitem[{Aifantis(1984)}]{Aifantis1984}
Aifantis, E.~C., 1984. {On the Microstructural Origin of Certain Inelastic
  Models}. Journal of Engineering Materials and Technology 106~(4), 326.

\bibitem[{Ashby(1970)}]{Ashby1970}
Ashby, M.~F., 1970. {The deformation of plastically non-homogeneous materials}.
  Philosophical Magazine 21~(170), 399--424.

\bibitem[{Bagchi and Evans(1996)}]{Bagchi1996}
Bagchi, A., Evans, A.~G., 1996. {The Mechanics and Physics of Thin-Film
  Decohesion and Its Measurement}. Interface Science 3, 169--193.

\bibitem[{Brinckmann and Siegmund(2008)}]{Brinckmann2008}
Brinckmann, S., Siegmund, T., 2008. {Computations of fatigue crack growth with
  strain gradient plasticity and an irreversible cohesive zone model}.
  Engineering Fracture Mechanics 75~(8), 2276--2294.

\bibitem[{Brown and Stobbs(1976)}]{Brown1976}
Brown, L.~M., Stobbs, W.~M., 1976. {The work-hardening of copper-silica V.
  Equilibrium plastic relaxation by secondary dislocations}. Philosophical
  Magazine 34~(3), 351--372.

\bibitem[{Chakravarthy and Curtin(2010)}]{Chakravarthy2010}
Chakravarthy, S.~S., Curtin, W.~A., 2010. {Origin of plasticity length-scale
  effects in fracture}. Physical Review Letters 105~(11), 1--4.

\bibitem[{Chen et~al.(1999)Chen, Wei, Huang, Hutchinson, and Hwang}]{Chen1999}
Chen, J.~Y., Wei, Y., Huang, Y., Hutchinson, J.~W., Hwang, K.~C., 1999. {The
  crack tip fields in strain gradient plasticity: the asymptotic and numerical
  analyses}. Engineering Fracture Mechanics 64~(5), 625--648.

\bibitem[{Cottrell(1964)}]{Cottrell1964}
Cottrell, A., 1964. {The Mechanical Properties of Materials}. Wiley, Oxford.

\bibitem[{Dahlberg and Faleskog(2013)}]{Dahlberg2013}
Dahlberg, C.~F., Faleskog, J., 2013. {An improved strain gradient plasticity
  formulation with energetic interfaces: Theory and a fully implicit finite
  element formulation}. Computational Mechanics 51~(5), 641--659.

\bibitem[{Danas et~al.(2012)Danas, Deshpande, and Fleck}]{Danas2012c}
Danas, K., Deshpande, V.~S., Fleck, N.~A., 2012. {Size effects in the conical
  indentation of an elasto-plastic solid}. Journal of the Mechanics and Physics
  of Solids 60~(9), 1605--1625.

\bibitem[{del Busto et~al.(2017)del Busto, Beteg{\'{o}}n, and
  Mart{\'{i}}nez-Pa{\~{n}}eda}]{EFM2017}
del Busto, S., Beteg{\'{o}}n, C., Mart{\'{i}}nez-Pa{\~{n}}eda, E., 2017. {A
  cohesive zone framework for environmentally assisted fatigue}. Engineering
  Fracture Mechanics 185, 210--226.

\bibitem[{Elssner et~al.(1994)Elssner, Korn, and R{\"{u}}hle}]{Elssner1994}
Elssner, G., Korn, D., R{\"{u}}hle, M., 1994. {The influence of interface
  impurities on fracture energy of UHV diffusion bonded metal-ceramic
  bicrystals}. Scripta Metallurgica et Materiala 31~(8), 1037--1042.

\bibitem[{Fleck and Hutchinson(1993)}]{Fleck1993}
Fleck, N.~A., Hutchinson, J.~W., 1993. {A phenomenological theory for strain
  gradient effects in plasticity}. Journal of the Mechanics and Physics of
  Solids 41~(12), 1825--1857.

\bibitem[{Fleck and Hutchinson(1997)}]{Fleck1997}
Fleck, N.~A., Hutchinson, J.~W., 1997. {Strain gradient plasticity}. Advances
  in Applied Mechanics 33, 295--361.

\bibitem[{Fleck and Hutchinson(2001)}]{Fleck2001}
Fleck, N.~A., Hutchinson, J.~W., 2001. {A reformulation of strain gradient
  plasticity}. Journal of the Mechanics and Physics of Solids 49~(10),
  2245--2271.

\bibitem[{Fleck et~al.(1994)Fleck, Muller, Ashby, and Hutchinson}]{Fleck1994}
Fleck, N.~A., Muller, G.~M., Ashby, M.~F., Hutchinson, J.~W., 1994. {Strain
  gradient plasticity: Theory and Experiment}. Acta Metallurgica et Materialia
  42~(2), 475--487.

\bibitem[{Fleck and Willis(2009{\natexlab{a}})}]{Fleck2009}
Fleck, N.~A., Willis, J.~R., 2009{\natexlab{a}}. {A mathematical basis for
  strain-gradient plasticity theory. Part II: Tensorial plastic multiplier}.
  Journal of the Mechanics and Physics of Solids 57~(7), 1045--1057.

\bibitem[{Fleck and Willis(2009{\natexlab{b}})}]{Fleck2009b}
Fleck, N.~A., Willis, J.~R., 2009{\natexlab{b}}. {A mathematical basis for
  strain-gradient plasticity theory—Part I: Scalar plastic multiplier}.
  Journal of the Mechanics and Physics of Solids 57, 161--177.

\bibitem[{Gao et~al.(1999)Gao, Hang, Nix, and Hutchinson}]{Gao1999}
Gao, H., Hang, Y., Nix, W.~D., Hutchinson, J.~W., 1999. {Mechanism-based strain
  gradient plasticity - I. Theory}. Journal of the Mechanics and Physics of
  Solids 47~(6), 1239--1263.

\bibitem[{Gudmundson(2004)}]{Gudmundson2004}
Gudmundson, P., 2004. {A unified treatment of strain gradient plasticity}.
  Journal of the Mechanics and Physics of Solids 52~(6), 1379--1406.

\bibitem[{Gurtin and Anand(2005)}]{Gurtin2005}
Gurtin, M.~E., Anand, L., 2005. {A theory of strain-gradient plasticity for
  isotropic, plastically irrotational materials. Part I: Small deformations}.
  International Journal of the Mechanics and Physics of Solids 53, 1624--1649.

\bibitem[{Hancock and Mackenzie(1976)}]{Hancock1976}
Hancock, J.~W., Mackenzie, A.~C., 1976. {On the mechanisms of ductile failure
  in high-strength steels subjected to multi-axial stress-states}. Journal of
  the Mechanics and Physics of Solids 24~(2-3), 147--160.

\bibitem[{Huang et~al.(2004)Huang, Qu, Hwang, Li, Gao, Huang, Qu, Hwang, Li,
  and Gao}]{Huang2004a}
Huang, Y., Qu, S., Hwang, K.~C., Li, M., Gao, H., Huang, Y., Qu, S., Hwang,
  K.~C., Li, M., Gao, H., 2004. {A conventional theory of mechanism-based
  strain gradient plasticity}. International Journal of Plasticity 20~(4-5),
  753--782.

\bibitem[{Jiang et~al.(2001)Jiang, Huang, Zhuang, and Hwang}]{Jiang2001}
Jiang, H., Huang, Y., Zhuang, Z., Hwang, K.~C., 2001. {Fracture in
  mechanism-based strain gradient plasticity}. Journal of the Mechanics and
  Physics of Solids 49~(5), 979--993.

\bibitem[{Jiang et~al.(2010)Jiang, Wei, Smith, Hutchinson, and
  Evans}]{Jiang2010}
Jiang, Y., Wei, Y., Smith, J.~R., Hutchinson, J.~W., Evans, A.~G., 2010. {First
  principles based predictions of the toughness of a metal/oxide interface}.
  International Journal of Materials Research 101, 1--8.

\bibitem[{Kanninen and Popelar(1985)}]{Kanninen1985}
Kanninen, M.~F., Popelar, C.~H., 1985. {Advanced Fracture Mechanics}. Oxford
  University Press.

\bibitem[{Komaragiri et~al.(2008)Komaragiri, Agnew, Gangloff, and
  Begley}]{Komaragiri2008}
Komaragiri, U., Agnew, S.~R., Gangloff, R.~P., Begley, M.~R., 2008. {The role
  of macroscopic hardening and individual length-scales on crack tip stress
  elevation from phenomenological strain gradient plasticity}. Journal of the
  Mechanics and Physics of Solids 56~(12), 3527--3540.

\bibitem[{Korn et~al.(2002)Korn, Elssner, Cannon, and Ruhle}]{Korn2002}
Korn, D., Elssner, G., Cannon, R.~M., Ruhle, M., 2002. {Fracture properties of
  interfacially doped Nb-A12O3 bicrystals: I, fracture characteristics}. Acta
  Materialia 50~(15), 3881--3901.

\bibitem[{Mart{\'{i}}nez-Pa{\~{n}}eda and Beteg{\'{o}}n(2015)}]{IJSS2015}
Mart{\'{i}}nez-Pa{\~{n}}eda, E., Beteg{\'{o}}n, C., 2015. {Modeling damage and
  fracture within strain-gradient plasticity}. International Journal of Solids
  and Structures 59, 208--215.

\bibitem[{Mart{\'{i}}nez-Pa{\~{n}}eda
  et~al.(2016{\natexlab{a}})Mart{\'{i}}nez-Pa{\~{n}}eda, del Busto, Niordson,
  and Beteg{\'{o}}n}]{IJHE2016}
Mart{\'{i}}nez-Pa{\~{n}}eda, E., del Busto, S., Niordson, C.~F., Beteg{\'{o}}n,
  C., 2016{\natexlab{a}}. {Strain gradient plasticity modeling of hydrogen
  diffusion to the crack tip}. International Journal of Hydrogen Energy
  41~(24), 10265--10274.

\bibitem[{Mart{\'{i}}nez-Pa{\~{n}}eda and Fleck(2018)}]{JAM2018}
Mart{\'{i}}nez-Pa{\~{n}}eda, E., Fleck, N.~A., 2018. {Crack growth resistance
  in metallic alloys: the role of isotropic versus kinematic hardening}.
  Journal of Applied Mechanics 85, 11002 (6 pages).

\bibitem[{Mart{\'{i}}nez-Pa{\~{n}}eda and Fleck(2019)}]{EJMaS2019}
Mart{\'{i}}nez-Pa{\~{n}}eda, E., Fleck, N.~A., 2019. {Mode I crack tip fields:
  strain gradient plasticity theory versus J2 flow theory}. European Journal of
  Mechanics, A/Solids (in press).

\bibitem[{Mart{\'{i}}nez-Pa{\~{n}}eda and Niordson(2016)}]{IJP2016}
Mart{\'{i}}nez-Pa{\~{n}}eda, E., Niordson, C.~F., 2016. {On fracture in finite
  strain gradient plasticity}. International Journal of Plasticity 80,
  154--167.

\bibitem[{Mart{\'{i}}nez-Pa{\~{n}}eda
  et~al.(2016{\natexlab{b}})Mart{\'{i}}nez-Pa{\~{n}}eda, Niordson, and
  Gangloff}]{AM2016}
Mart{\'{i}}nez-Pa{\~{n}}eda, E., Niordson, C.~F., Gangloff, R.~P.,
  2016{\natexlab{b}}. {Strain gradient plasticity-based modeling of hydrogen
  environment assisted cracking}. Acta Materialia 117, 321--332.

\bibitem[{McClintock(1968)}]{McClintock1968}
McClintock, F.~A., 1968. {A Criterion for Ductile Fracture by the Growth of
  Holes}. Journal of Applied Mechanics 35~(2), 363.

\bibitem[{Nielsen and Niordson(2014)}]{Nielsen2014}
Nielsen, K.~L., Niordson, C.~F., 2014. {A numerical basis for strain-gradient
  plasticity theory: Rate-independent and rate-dependent formulations}. Journal
  of the Mechanics and Physics of Solids 63~(1), 113--127.

\bibitem[{Nix and Gao(1998)}]{Nix1998}
Nix, W.~D., Gao, H.~J., 1998. {Indentation size effects in crystalline
  materials: A law for strain gradient plasticity}. Journal of the Mechanics
  and Physics of Solids 46~(3), 411--425.

\bibitem[{Nye(1953)}]{Nye1953}
Nye, J.~F., 1953. {Some geometrical relations in dislocated crystals}. Acta
  Metallurgica 1~(2), 153--162.

\bibitem[{Panteghini and Bardella(2016)}]{Panteghini2016}
Panteghini, A., Bardella, L., 2016. {On the Finite Element implementation of
  higher-order gradient plasticity, with focus on theories based on plastic
  distortion incompatibility}. Computer Methods in Applied Mechanics and
  Engineering 310, 840--865.

\bibitem[{Papazafeiropoulos et~al.(2017)Papazafeiropoulos,
  Mu{\~{n}}iz-Calvente, and Mart{\'{i}}nez-Pa{\~{n}}eda}]{AES2017}
Papazafeiropoulos, G., Mu{\~{n}}iz-Calvente, M., Mart{\'{i}}nez-Pa{\~{n}}eda,
  E., 2017. {Abaqus2Matlab: A suitable tool for finite element
  post-processing}. Advances in Engineering Software 105, 9--16.

\bibitem[{Poole et~al.(1996)Poole, Ashby, and Fleck}]{Poole1996}
Poole, W.~J., Ashby, M.~F., Fleck, N.~A., 1996. {Micro-hardness of annealed and
  work-hardened copper polycrystals}. Scripta Materialia 34~(4), 559--564.

\bibitem[{Pribe et~al.(2019)Pribe, Siegmund, Tomar, and Kruzic}]{Pribe2019}
Pribe, J.~D., Siegmund, T., Tomar, V., Kruzic, J.~J., 2019. {Plastic strain
  gradients and transient fatigue crack growth: a computational study}.
  International Journal of Fatigue 120, 283--293.

\bibitem[{Rice(1968)}]{Rice1968a}
Rice, J.~R., 1968. {Mathematical Analysis in the Mechanics of Fracture}.
  Mathematical Fundamentals 2~(B2), 191--311.

\bibitem[{Rice and Tracey(1969)}]{Rice1969}
Rice, J.~R., Tracey, D.~M., 1969. {On the ductile enlargement of voids in
  triaxial stress fields}. Journal of the Mechanics and Physics of Solids
  17~(3), 201--217.

\bibitem[{Segurado and LLorca(2004)}]{Segurado2004}
Segurado, J., LLorca, J., 2004. {A new three-dimensional interface finite
  element to simulate fracture in composites}. International Journal of Solids
  and Structures 41~(11-12), 2977--2993.

\bibitem[{Seiler et~al.(2016)Seiler, Siegmund, Zhang, Tomar, and
  Kruzic}]{Seiler2016}
Seiler, P.~E., Siegmund, T., Zhang, Y., Tomar, V., Kruzic, J.~J., 2016.
  {Stationary and propagating cracks in a strain gradient visco-plastic solid}.
  International Journal of Fracture 202~(1), 111--125.

\bibitem[{Sevillano(2001)}]{Sevillano2001}
Sevillano, J.~G., 2001. {The effective threshold for fatigue crack propagation:
  A plastic size effect?} Scripta Materialia 44~(11), 2661--2665.

\bibitem[{Shu and Fleck(1999)}]{Shu1999}
Shu, J.~Y., Fleck, N.~A., 1999. {Strain gradient crystal plasticity:
  size-dependent deformation of bicrystals}. Journal of the Mechanics and
  Physics of Solids 47~(2), 297--324.

\bibitem[{Stelmashenko et~al.(1993)Stelmashenko, Walls, Brown, and
  Milman}]{Stelmashenko1993}
Stelmashenko, N.~A., Walls, M.~G., Brown, L.~M., Milman, Y.~V., 1993.
  {Microindentations on W and Mo oriented single crystals: An STM study}. Acta
  Metallurgica Et Materialia 41~(10), 2855--2865.

\bibitem[{St{\"{o}}lken and Evans(1998)}]{Stolken1998}
St{\"{o}}lken, J.~S., Evans, A.~G., 1998. {A microbend test method for
  measuring the plasticity length scale}. Acta Materialia 46~(14), 5109--5115.

\bibitem[{Suo et~al.(1993)Suo, Shih, and Varias}]{Suo1993}
Suo, Z., Shih, C.~F., Varias, A.~G., 1993. {A theory for cleavage cracking in
  the presence of plastic flow}. Acta Metallurgica Et Materialia 41~(5),
  1551--1557.

\bibitem[{Tvergaard(1976)}]{Tvergaard1976}
Tvergaard, V., 1976. {Effect of thickness inhomogeneities in internally
  pressurized elastic-plastic spherical shells}. Journal of the Mechanics and
  Physics of Solids 24~(5), 291--304.

\bibitem[{Tvergaard and Hutchinson(1992)}]{Tvergaard1992}
Tvergaard, V., Hutchinson, J.~W., 1992. {The relation between crack growth
  resistance and fracture process parameters in elastic-plastic solids}.
  Journal of the Mechanics and Physics of Solids 40~(6), 1377--1397.

\bibitem[{Tvergaard and Niordson(2008)}]{Tvergaard2008}
Tvergaard, V., Niordson, C.~F., 2008. {Size effects at a crack-tip interacting
  with a number of voids}. Philosophical Magazine 88~(30-32), 3827--3840.

\bibitem[{Wei and Hutchinson(1997)}]{Wei1997}
Wei, Y., Hutchinson, J.~W., 1997. {Steady-state crack growth and work of
  fracture for solids characterized by strain gradient plasticity}. Journal of
  the Mechanics and Physics of Solids 45~(8), 1253--1273.

\bibitem[{Wei et~al.(2004)Wei, Qiu, and Hwang}]{Wei2004}
Wei, Y., Qiu, X., Hwang, K.~C., 2004. {Steady-state crack growth and fracture
  work based on the theory of mechanism-based strain gradient plasticity}.
  Engineering Fracture Mechanics 71~(1), 107--125.

\end{thebibliography}


\end{document}